\documentstyle[aps,pre,psfig,eqsecnum]{revtex}

\begin{document}
\draft

\twocolumn[\hsize\textwidth\columnwidth\hsize\csname @twocolumnfalse\endcsname

\title{Multiple light scattering in anisotropic random media}

\author{Holger Stark\thanks{permanent address:
Institut f\"ur Theoretische 
und Angewandte Physik, Universit\"at Stuttgart, Pfaffenwaldring 57, 70550 
Stuttgart, Germany} and Tom C. Lubensky}
\address{Department of Physics \& Astronomy, University of Pennsylvania,
Philadelphia, PA 19104, USA}

\maketitle

\begin{abstract}
In the last decade Diffusing Wave Spectroscopy (DWS) has emerged as
a powerful tool to study turbid media. In this article we develop the
formalism to describe light diffusion in general anisotropic turbid media. We
give explicit formulas to calculate the diffusion tensor and the
dynamic absorption coefficient, measured in DWS experiments.
We apply our theory to uniaxial systems, namely nematic liquid crystals,
where light is scattered from thermal fluctuations of the local optical axis,
called director.
We perform a detailed analysis of the two essential diffusion constants,
parallel and perpendicular to the director, in terms of Frank elastic 
constants, dielectric anisotropy, and applied magnetic field.
We also point out the relevance of our results to different liquid crystalline
systems, such as discotic nematics, smectic-$A$ phases, and polymer liquid
crystals. Finally, we show that the dynamic absorption coefficient is the 
angular average over the inverse viscosity, which governs the dynamics of
director fluctuations.
\end{abstract}


\vskip2pc]

\narrowtext

\section{Introduction} \label{sec intro}

Dynamic light scattering (DLS) is one of our most powerful probes
\cite{Berne76,Clark70,Dhont83} of the
dynamics of materials like simple liquids, complex
fluids, and liquid crystals.  In typical
experiments, light incident on the sample scatters once, and its intensity
is measured at a detector.  Motion in the sample, or more generally
fluctuations in the local dielectric constant, induce changes in the phase
of scattered light, which give rise to temporal fluctuations 
of the light intensity measured at the detector.  Such
experiments probe length scales of order the inverse scattering wave vector
$q^{-1}$ and hence time scales of order $(q v )^{-1}$ where $v$ is a
typical velocity.

There are many materials such as colloids, emulsions, foams,
and some liquid crystals that
scatter light so strongly that the traditional single scattering analysis
of DLS does not apply.  In these materials, light undergoes many scattering
events before leaving the sample, and the transport of light energy is
diffusive rather than ballistic. The study of light transport in random or
turbid media dates back to radiative transfer theory, first introduced as
early as 1905 by Schuster\cite{Schuster05}. These systems are characterized
by a scattering mean-free-path $l$, measuring the average distance a photon 
travels
before scattering, and a transport mean-free-path $l^* = l/\langle 1 -
\cos\vartheta_s\rangle$, measuring the distance beyond which the direction of
propagation is randomized, where $\vartheta_s$ is the scattering angle and the
angular bracket denotes an average over direction weighted by the
differential scattering cross section.  For distances greater than $l^*$, 
light energy transport is described by a diffusion equation with scalar 
diffusion constant $D = {\overline c} l^*/3$, where ${\overline c}$ is the 
speed of light in the medium.

There has been a resurgence of interest in light transport in turbid 
and random media\cite{Sheng90} 
because of its close connection to the problem of Anderson
localization\cite{Anderson58} 
and electron transport in disordered systems\cite{Abrahams79,Lee85}
and because of the development of Diffusing
Wave Spectroscopy (DWS)\cite{Maret87,Rosenbluh87b,Pine88,Pine90,Golubentsev84,Stephen88,MacKintosh89}, which permits useful information to be
extracted from dynamic correlations of multiply scattered light. 
Coherent backscattering, a manifestation of weak as opposed to strong or
true Anderson localization, has been observed in a number of
experiments\cite{Kuga84,Albada85,Wolf85,Etemad86,Kaveh86,Etemad87,Rosenbluh87a}
 and discussed in a number of theoretical papers
\cite{Golubentsev84,Akkermans85,Akkermans86,Stephen86,Mark88,Akkermans88,MacKintosh88}.  
DWS has opened up a
whole new field of study. It provides heretofore unobtainable information
about the the dynamics of turbid media, including dense 
colloids \cite{Pine90}, sheared suspensions \cite{Wu90}
emulsions \cite{Gang94}, and foams \cite{Durian91}.  
Because intensity variations measured at the
detector arise from phase shifts distributed over many scattering events,
DWS detects dynamic phenomena at much shorter time scales than normal DLS.
This has permitted the measurement of hydrodynamic interaction 
contributions to the
diffusion constant of colloidal particles \cite{Kao93} and the measurement
of shape fluctuation modes in tense emulsion droplets \cite{Gang94}. 
Photon diffusion and DWS have also found application in imaging of objects
such a tumors in 
human tissue \cite{Yodh95}.

With few exceptions \cite{Vlasov88,Vithana93,Ramaswamy94,Tiggelen95}, 
both theory and experiment have focussed
on diffusive transport and DWS in isotropic systems.  There are, however,
many turbid materials such as conventional thermotropic and lyotropic liquid 
crystals, liquid crystalline colloids \cite{Fraden95,Kirchhoff96} and 
emulsions, and also muscle tissues that are anisotropic. 
This paper will develop a general treatment of diffusive light
transport and DWS in anisotropic media with particular applications to
nematic liquid crystals.  Though its inspiration is recent experimental work on
coherent backscatter \cite{Vlasov88,Vithana93}, its purpose is to broaden the 
class of materials to which DWS and its offshoot applications like imaging 
can be applied. A preliminary account of this 
work and experiments on multiple scattering in liquid crystals to which it
applies were reported in references \cite{Stark96a,Kao96}.  An alternative
derivation of the results reported here and a more detailed account of
experiments appear in Ref. \cite{Jester96}. A similar treatment of
diffusive light transport and DWS was developed by Tiggelen, Maynard,
and Heiderich \cite{Tiggelen96}.

Anisotropic media differ from isotropic media in two important ways: (1)
The speed of light depends in general on both the polarization and direction
of light propagation relative to
anisotropy axes, and (2) scattering cross sections depend not 
only on the relative direction of incoming and scattered light rays but also 
on their direction relative to anisotropy axes. 
Diffusive transport in optically active isotropic media with light speeds 
depending on the state of circular polarization has been studied
\cite{MacKintosh88}. Tiggelen has investigated anisotropic light 
diffusion induced by a magnetic field $H$ through a series expansion in 
$H$ \cite{Tiggelen95}.
To our knowledge, however, no thorough treatment of multiple light scattering 
in optically anisotropic media has been published. For electronic
systems anisotropic diffusion has been studied both theoretically and
experimentally in the context of localization \cite{Woelfle84,Bishop84}.

Optical anisotropy leads to anisotropic diffusive light transport.  The
equation governing the electric-field autocorrelation function
$W(\bbox{R},T,t) = \langle \, \bbox{E}(\bbox{R}, T +t/2 ) \cdot 
\bbox{\varepsilon}_0 \bbox{E}(\bbox{R}, T-t/2 ) \,\rangle$ (
$\bbox{\varepsilon}_0$ denotes the dielectric tensor) is
\begin{equation}
\Bigl[{\partial\over \partial T} - \bbox{\nabla}\cdot\bbox{D}\bbox{\nabla} + 
\mu (t)\Bigl] \, W(\bbox{R},T,t) = \varrho(\bbox{R},T) \enspace ,
\end{equation}
where $\bbox{D}$ stands for the anisotropic diffusion tensor. The quantity
$\mu(t)$ is the dynamic absorption coefficient measured in DWS experiments
\cite{Maret87,Rosenbluh87b,Pine88,Pine90,Golubentsev84,Stephen88,MacKintosh89}.
It results from an average of short-time dynamic correlations over angle and 
polarization. We will provide explicit formulas for $\bbox{D}$ and $\mu(t)$ 
for general anisotropic systems and then concentrate on nematic liquid 
crystals as one example of a uniaxial system. With the preferred axis
along the unit vector $\bbox{n}$, the diffusion tensor $\bbox{D}$ reduces to 
$\bbox{D} = D_{\perp}\bbox{1} + (D_{\|} -D_{\perp}) \bbox{n} \otimes \bbox{n}$
 where $\bbox{1}$ is the unit element. 
The dynamic absorption coefficient $\mu(t)$ will turn out to be the angular 
average of an inverse viscosity.


In this paper, we will restrict ourselves to the weak-scattering limit, and
we will treat multiple scattering via the Bethe-Salpeter equation
in Sec. III. 
In isotropic lossless systems, the diffusion equation can be
obtained exactly from the Bethe-Salpeter equation by considering only
modes associated with the isotropic and ``vector" spherical harmonics
$Y_{00}$ and $Y_{1m}$.  In anisotropic systems, all spherical harmonics
couple, and the calculation of diffusion coefficients involves the
inversion of an infinite dimensional matrix, which can only be 
accomplished approximately.  We will, therefore,  be content with a formal 
expression for the diffusion tensor in general anisotropic media. 
We will, however,
introduce a sequence of approximations to obtain numerical values for the
diffusion constants in nematic liquid crystals.  Fortunately, the first term in
this sequence undergoes only a very small modification in going to the second 
in this sequence.

Nematic liquid crystals present a difficulty that is not generic to
anisotropic systems.  Light scattering is from fluctuations in the
direction of the principal axis of the dielectric tensor, which is
parallel to the local Frank director $\bbox{n}(\bbox{r})$.  
Fluctuations in $\bbox{n}(\bbox{r})$
diverge as $q^{-2}$ at small wave number $q$ in the absence of an external
aligning magnetic field $H$.  This divergence leads to a vanishing scattering
mean-free-path for extraordinary-to-extraordinary scattering in the limit
$H \rightarrow 0$. The diffusion constants are nonetheless well defined and
non-zero. For, if $l$ tends to zero, scattering takes place almost entirely 
in the forward direction. Thus, light has to undergo a large number of
scattering events before directional information is lost, and, as a result,
$l^*$ is finite.


The outline of this paper is as follows.  In Sec. II, we review light
propagation including the one-particle Green function in homogeneous 
anisotropic media. In Sec. III, we treat diffusive transport of light in 
general anisotropic random media. We introduce the structure factor 
$\bbox{B}^{\omega}(\bbox{r},t)$ to describe fluctuations in the
dielectric tensor, discuss electric-field autocorrelation functions and
their meaning, and relate them to the averaged two-particle 
Green function. We, then, discuss the one- and two-particle Green 
functions in the weak-scattering limit and derive the diffusion 
equation for light transport from the Bethe-Salpeter equation. We introduce
the approximation scheme for the diffusion tensor and look at the isotropic
limit of our theory. Comments on DWS close Sec.~III.
Sec. IV applies the general results of the preceding sections to nematic 
liquid crystals. A review of relevant properties of nematic liquid crystals 
and light propagation in uniaxial media is followed by an explanation of
dielectric tensor fluctuations in nematics.
Finally, we discuss diffusive light transport and DWS in nematics.  In
particular, we provide explicit numerical calculations of the diffusion
coefficients $D_{\parallel}$ and $D_{\perp}$  as a function of Frank
elastic constants, dielectric anisotropy, and external magnetic field, and
point out their relevance for different liquid crystalline systems, such as 
discotic nematics, smectic-$A$ phases, and polymer liquid crystals.
The numerical calculations are summarized in Figures \ref{fig5}-\ref{fig9}.
They are in excellent agreement with recent experiments on the nematic
compound 5CB by Jester, Kao, and Yodh \cite{Kao96,Jester96}. At the end we
address the dynamic absorption coefficient.

\section{Light propagation in a homogeneous medium with dielectric
anisotropy} \label{sec homo}

Light propagation in anisotropic dielectric media is more complicated than it
is in isotropic systems. In particular, the electric field is not always
transverse, and the speed of light depends on polarization and direction
of propagation. In this section we review light propagation in anisotropic
media. Following the work of Nelson and Lax \cite{Lax71},
we will introduce sets of polarization vectors
for the electric and dielectric field that will prove to be very useful
for our forthcoming considerations. We start with Maxwell's equations
\begin{equation}
\label{2.0}
 \begin{array}{rcl@{\quad}rcl}
  \text{div} \bbox{D} & = & 4\pi \varrho_{\text{ma\rule[-3mm]{0mm}{3mm}}} &
  \text{div} \bbox{B} & = & 0 \\ 
  \text{curl} \bbox{E} & = &\displaystyle - \frac{1}{c} \, 
  \frac{\partial \bbox{B}}{\partial t} &
  \text{curl} \bbox{H} & = & \displaystyle 
  \frac{4\pi}{c}\,\bbox{j}_{\text{ma}} +
  \frac{1}{c} \, \frac{\partial \bbox{D}}{\partial t} \,\, ,
 \end{array}
\end{equation}
where $\varrho_{\text{ma}}$ and $\bbox{j}_{\text{ma}}$ are, respectively,
the macroscopic charge and current densities. We concentrate on a
dielectric medium with
\begin{equation}
\label{2.01}
\bbox{D} = \bbox{\varepsilon}_0 \, \bbox{E} \quad \text{and} \quad
\bbox{B} = \bbox{H} \enspace,
\end{equation}
and we assume that the dielectric tensor $\bbox{\varepsilon}_0$ is real and 
does not depend on time. Then the energy-balance equation reads:
\begin{equation}
\label{2.02}
\frac{\partial}{\partial t} \, u + \text{div} \bbox{S} = \bbox{j}_{\text{ma}}
\cdot \bbox{E} \enspace,
\end{equation}
where we have introduced the energy density
\begin{equation}
\label{2.03}
u = \frac{1}{8\pi} \, (\bbox{E} \cdot \bbox{\varepsilon}_0 \bbox{E}
  + \bbox{H} \cdot \bbox{H} )
\end{equation}
and the Poynting vector
\begin{equation}
\label{2.04}
\bbox{S} = \frac{c}{4\pi} \, \bbox{E} \times \bbox{H} \enspace.
\end{equation}
Both quantities strongly vary in space and time. If we average over one
period of oscillation, we obtain their averaged values $\bar{u}$ and 
$\overline{\bbox{S}}$. In the following we use complex waves 
$\bbox{E}=\bbox{E}_0 e^{-i\omega t}$ and $\bbox{H}=\bbox{H}_0 e^{-i\omega t}$
whose averaged energy densities and Poynting vectors are given by
$\bar{u}= (\bbox{E}_0 \cdot \bbox{\varepsilon} \bbox{E}_0 + \bbox{H}_0
\cdot \bbox{H}_0)/(16\pi)$ and $\overline{\bbox{S}} =
c(\bbox{E}_0 \times \bbox{H}_0) / (8\pi)$.

For vanishing sources $\varrho_{\text{ma}}$ and $\bbox{j}_{\text{ma}}$, we 
obtain the homogeneous wave equation for the electric light field 
$\bbox{E}(\bbox{r},t)$:
\begin{equation}
\label{2.1}
[\text{curl}\,\text{curl} + 
\frac{\bbox{\varepsilon}_0}{c^2}\, \frac{\partial^2}{\partial t^2}] \, 
\bbox{E}(\bbox{r},t) = \bbox{0} \enspace.
\end{equation}
Note that all solutions of Eq.\ (\ref{2.1}) have to fulfill
the transversality condition for dielectric field waves, 
$\text{div}\bbox{\varepsilon}_{0} \bbox{E} =0$, unless the electric field
is static or linear in $t$. Introduction of the plane-wave ansatz
\begin{equation}
\label{2.2}
\bbox{E}(\bbox{r},t) = E_0 \bbox{e}(\hat{\bbox{k}})\,
\text{exp}[i(\bbox{k}\cdot \bbox{r} - \omega t)]
\end{equation}
in wave equation (\ref{2.1}) leads to a generalized eigenvalue problem
\begin{equation}
\label{2.3}
[\bbox{P}_{t}(\hat{\bbox{k}}) - \frac{1}{n^2}\, \bbox{\varepsilon}_0] \,
\bbox{e}(\hat{\bbox{k}}) = \bbox{0} \enspace,
\end{equation}
where
\begin{equation}
\frac{1}{n^2} = \frac{\omega^2}{c^2 k^2}
\end{equation}
and
\begin{equation}
\label{2.4}
\bbox{P}_{t}(\hat{\bbox{k}}) = \bbox{1}-\hat{\bbox{k}}\otimes
\hat{\bbox{k}}
\end{equation}
is the projection operator on the space perpendicular to the unit vector
$\hat{\bbox{k}}=\bbox{k}/k$. The symbol $\otimes$ stands for the tensor
product: $[\hat{\bbox{k}}\otimes \hat{\bbox{k}}]_{ij} = \hat{k}_i \hat{k}_j$.
The solutions to Eq.\ (\ref{2.3}) provide us with the characteristic light 
modes of the system
determined by the refractive index $n_i(\hat{\bbox{k}})= ck/\omega$ and
the polarization vector $\bbox{e}_{i}(\hat{\bbox{k}})$ of the electric field.
We define the polarization vector of the displacement field by 
\begin{equation}
\label{2.4b}
\bbox{d}^{\,i}(\hat{\bbox{k}}) = 
\bbox{\varepsilon}_0 \bbox{e}_{i}(\hat{\bbox{k}}) \enspace.
\end{equation}
For each direction $\hat{\bbox{k}}$, there exist two characteristic light 
modes with associated polarizations 
$\bbox{d}^{\,i}(\hat{\bbox{k}}) \perp \hat{\bbox{k}}$. 
When a plane wave with frequency $\omega$ enters the 
anisotropic medium, it splits up into the two characteristic modes which 
travel with different speeds $c/n_i(\hat{\bbox{k}})$ and wave numbers
$k=\omega n_i / c$. We also find a third solution with $n_3=\infty$ and
$\bbox{e}_3(\hat{\bbox{k}}) \| \hat{\bbox{k}}$, corresponding to a 
nonpropagating mode with $\omega=0$. It violates 
$\hat{\bbox{k}} \cdot \bbox{d}^3 =0$ but is necessary to construct the
complete Green function for wave equation (\ref{2.1}) (see below).

\begin{figure}
\centerline{\psfig{figure=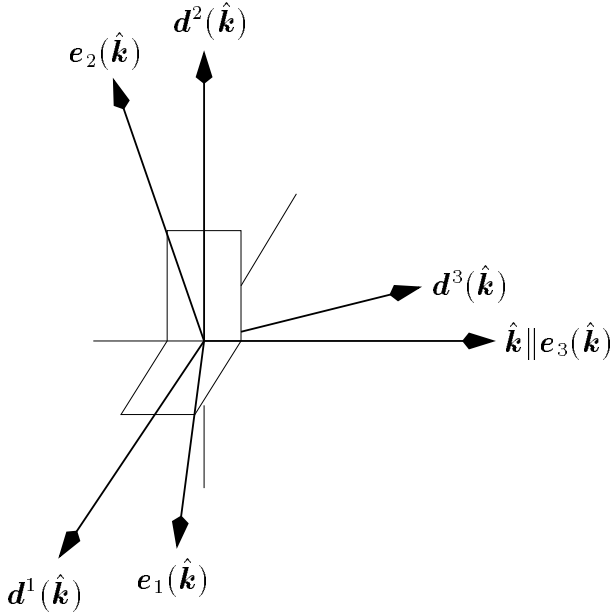}}

\vspace*{.5cm}

\caption[]{The polarization vectors $\bbox{e}_i(\hat{\bbox{k}})$ and 
$\bbox{d}^{\,i}(\hat{\bbox{k}})$ for a given propagation direction 
$\hat{\bbox{k}}$ in an anisotropic medium.}
\label{fig1}
\end{figure}

The two sets of polarization vectors fulfill the biorthogonality condition
\begin{equation}
\label{2.5}
\bbox{d}^{\,i}(\hat{\bbox{k}}) \cdot \bbox{e}_{j}(\hat{\bbox{k}}) =
\delta^i_j \quad(i,j=1,2,3)\enspace,
\end{equation}
i.~e., they are dual to each other like the basis vectors
of the real and reciprocal lattice in a crystal. We will use both of them as
convenient bases for our tensor quantities throughout the article. 
To prove
condition (\ref{2.5}) we notice that $\bbox{P}_{t}(\hat{\bbox{k}})$ and 
$\bbox{\varepsilon}_0$ are symmetric tensors and derive from Eq.\ (\ref{2.3})
the condition
\begin{equation}
\label{2.6}
\Bigl( \frac{1}{n_i^2} - \frac{1}{n_j^2} \Bigr) \, \bbox{e}_i \cdot 
\bbox{d}^{\,j}
= 0 \enspace,
\end{equation}
from which (\ref{2.5}) follows after an appropriate normalization.
The vectors $\bbox{d}^1$ and $\bbox{d}^2$ are perpendicular to 
$\hat{\bbox{k}}$. Then, again with Eq.\ (\ref{2.3}) and the biorthogonality
condition\ (\ref{2.5}), one can show that $\bbox{d}^1 \perp \bbox{d}^2$.
In Fig.\ \ref{fig1}, we
summarize the geometry for a given propagation direction $\hat{\bbox{k}}$.
%
Finally, we recall that in general the refractive indices are calculated from
Fresnel's equation \cite{Landau60}:
\begin{equation}
\label{2.7}
\sum_{i=1}^3 \frac{\bar{\varepsilon}_i \hat{k}_i^2}{n^2-\bar{\varepsilon}_i}
=0
\enspace,
\end{equation}
where $\bar{\varepsilon}_i$ stands for the principal dielectric constants, the
eigenvalues of $\bbox{\varepsilon}_0$, and $\hat{k}_i$ is the component of the 
unit vector $\hat{\bbox{k}}$ along the $i$th eigenvector of
$\bbox{\varepsilon}_0$. This equation is equivalent to
$\text{det}[\bbox{P}_{t}(\hat{\bbox{k}}) - \bbox{\varepsilon}_0/n^2]=0$.

Consider now $\bbox{E}_0 = E_0 \bbox{e}_i(\hat{\bbox{k}})$ and
$\bbox{H}_0$, the amplitude of the magnetic field wave, which are connected
via Maxwell's relations:
\begin{equation}
\label{2.7a}
\bbox{k} \times \bbox{E}_0 = \frac{\omega}{c} \, \bbox{H}_0 \quad \text{and}
\quad \bbox{k} \times \bbox{H}_0 = - \frac{\omega}{c} \, 
\bbox{\varepsilon}_0 \bbox{E}_0 \enspace.
\end{equation}
The second equation also follows from the first once Eq.\ (\ref{2.3}) is 
solved.
For the nonpropagating mode ($\omega = 0$), $\bbox{H}_0$, like $\bbox{E}_0$,
is parallel to $\hat{\bbox{k}}$. For the propagating modes, 
$\hspace*{-.3cm}$$\bbox{H}_0$ is
perpendicular to $\hat{\bbox{k}}$ and $\bbox{\varepsilon}_0 \bbox{E}_0$.
Using Eqs.\ (\ref{2.3}) and (\ref{2.7a}) one shows that electric and magnetic
fields carry the same amount of field energy ($
\bbox{H}_0 \cdot \bbox{H}_0^{\ast} = \bbox{E}_0 \cdot 
\bbox{\varepsilon}_0 \bbox{E}_0^{\ast}
$) and that the averaged energy density $\bar{u}$ of a light wave is
\begin{equation}
\label{2.7b}
\bar{u} = \frac{1}{8\pi} \, \bbox{E}_0 \cdot 
\bbox{\varepsilon}_0 \bbox{E}_0^{\ast} = \frac{1}{8\pi} \, |E_0|^2
\enspace.
\end{equation}
Here the meaning of the scalar $E_0$ becomes clear. It is not the magnitude
of the electric field, because $\bbox{e}_i(\hat{\bbox{k}})$ is not a unit 
vector,
instead it basically stands for the square root of the energy density of the 
light mode. The poynting vector $\bbox{S}$ does not generally point along
$\hat{\bbox{k}}$. Its projection on $\hat{\bbox{k}}$ however fulfills
the relation
\begin{equation}
\label{2.7c}
\hat{\bbox{k}} \cdot \overline{\bbox{S}} = \frac{c}{n} \, 
\bar{u} \enspace ,
\end{equation}
which is familiar for isotropic systems.

Finally, we calculate the Green function
$\bbox{G}_0(\bbox{r}-\bbox{r'},t-t')$ for the wave equation (\ref{2.1}),
which has a
source term proportional to the time derivative of $\bbox{j}_{\text{ma}}$.
%
%
In Fourier space we have
\begin{equation}
\label{2.9}
\bigl[ k^2 \bbox{P}_t(\hat{\bbox{k}}) - \frac{\omega^2}{c^2}\,
\bbox{\varepsilon}_0 \bigr] \, \bbox{G}_0(\bbox{k},\omega) = -\bbox{1} 
\enspace.
\end{equation}
We expand $\bbox{G}_0(\bbox{k},\omega)$ in the basis
$\{ \bbox{e}_i(\hat{\bbox{k}}) \otimes \bbox{e}_j(\hat{\bbox{k}}) \}$ and 
calculate its components by multiplying the last equation, respectively,
from left and right with $\bbox{e}_l(\hat{\bbox{k}})$ and
$\bbox{d}^k(\hat{\bbox{k}})$. Using the eigenvalue equation (\ref{2.3}) and
the algebra for the polarization vectors, described above, we find that 
$\bbox{G}_0(\bbox{k},\omega)$ is diagonal:
\begin{eqnarray}
\bbox{G}_0(\bbox{k},\omega) & = & \sum_{\alpha=1}^{2}\,
\left[ \frac{\omega^2}{c^2} - \frac{k^2}{n_{\alpha}^2(\hat{\bbox{k}})}
\right]^{-1} \, \bbox{e}_{\alpha}(\hat{\bbox{k}}) \otimes 
\bbox{e}_{\alpha}(\hat{\bbox{k}}) \nonumber \\
\label{2.10}
 & & \qquad \quad + \, \frac{c^2}{\omega^2}\, \bbox{e}_{3}(\hat{\bbox{k}}) 
 \otimes \bbox{e}_{3}(\hat{\bbox{k}}) \enspace.
\end{eqnarray}
The first term on the right-hand side represents the propagating part of
the Green function. We will always, throughout this article, indicate it
by Greek indices. It acts on the transverse part of the source term.
Nelson and Lax \cite{Lax73} have calculated an expression for its far field
in coordinate space, which reduces to a spherical wave in an isotropic system.
The second term takes into account the longitudinal part of the source. 
In coordinate space, it is just a non-propagating delta function. 
As we go on, it will become clear that we 
can neglect it within our approximation outlined in the next section. We will
therefore skip it when we represent tensors through their components.
We will also see that for all the tensors involved we can neglect the 
non-diagonal parts. Thus, whenever we use a Greek superscript or
subscript $\alpha$ it refers, respectively, to the basis "vectors"
$\bbox{e}_{\alpha}(\hat{\bbox{k}}) \otimes \bbox{e}_{\alpha}(\hat{\bbox{k}})$
or $\bbox{d}^{\alpha}(\hat{\bbox{k}}) \otimes \bbox{d}^{\alpha}(\hat{\bbox{k}})
$ ($\alpha=1,2$).
The last quantity we need is the inverse Green function 
$[\bbox{G}_0(\bbox{k},\omega)]^{-1}$, which we conveniently represent as
\begin{eqnarray}
[\bbox{G}_0(\bbox{k},\omega)]^{-1} & = & \sum_{\alpha=1}^{2}\,
\Bigl[ \frac{\omega^2}{c^2} - \frac{k^2}{n_{\alpha}^2(\hat{\bbox{k}})} \Bigr]
 \, \bbox{d}^{\alpha}(\hat{\bbox{k}}) \otimes \bbox{d}^{\alpha}(\hat{\bbox{k}})
\nonumber \\
\label{2.11}
 & & \qquad \quad + \, \frac{\omega^2}{c^2}\, 
 \bbox{d}^{3}(\hat{\bbox{k}}) \otimes \bbox{d}^{3}(\hat{\bbox{k}})
\end{eqnarray}
by realizing that the unit tensor can be written as
\begin{equation}
\label{2.12}
\bbox{1} = \sum_{ij} \delta^i_j \,\bbox{e}_{i}(\hat{\bbox{k}}) \otimes
 \bbox{d}^{j}(\hat{\bbox{k}}) = \sum_{ij} \delta_i^j \, 
 \bbox{d}^{i}(\hat{\bbox{k}}) \otimes \bbox{e}_{j}(\hat{\bbox{k}}) \enspace.
\end{equation}

\section{Diffusive light transport in anisotropic random media} 
\label{sec rand}


We now consider a dielectric medium in which the dielectric tensor possesses 
a randomly fluctuating part $\delta \bbox{\varepsilon}(\bbox{r},t)$ in addition
to the homogeneous term $\bbox{\varepsilon}_0$. We treat 
$\delta \bbox{\varepsilon}(\bbox{r},t)$ as a Gaussian random variable with
variance characterized by
\begin{equation}
\label{3.1}
\bbox{B}^{\omega}(\bbox{r},t):= \frac{\omega^4}{c^4}\,
\langle \, \delta \bbox{\varepsilon}(\bbox{r},t) \otimes 
\delta \bbox{\varepsilon}(\bbox{0},0) \, \rangle^{(N)} \enspace ,
\end{equation}
where $\omega$ is the frequency of light. The superscript $(N)$ means that we 
interchange the second and third index in the tensor product
$\delta \bbox{\varepsilon} \otimes \delta \bbox{\varepsilon}$ to
define $\bbox{B}^{\omega}$: $[\bbox{B}^{\omega}]_{ijkl} \propto \langle \,
\delta \varepsilon_{ik} \, \delta \varepsilon_{jl} \, \rangle$. We call 
$\bbox{B}^{\omega}(\bbox{r},t)$ the structure factor of the
system. It is measured in single light scattering experiments
\cite{Berne76} and contains information about the elastic and dynamic
properties of a system.

The electric light field $\bbox{E}(\bbox{r},t)$ obeys the homogeneous wave 
equation
\begin{equation}
\label{3.2}
[\text{curl}\, \text{curl}  +
(\bbox{\varepsilon}_0 + \delta \bbox{\varepsilon}(\bbox{r},t)) \,
\frac{1}{c^2}\,
\frac{\partial^2}{\partial t^2} \, ] \, \bbox{E}(\bbox{r},t) = \bbox{0} 
\enspace,
\end{equation}
where we have used an adiabatic approximation to pull 
$\delta \bbox{\varepsilon}(\bbox{r},t)$ in front of the time derivatives.
It is valid if $\delta \bbox{\varepsilon}(\bbox{r},t)$ varies on
time scales much longer than the time period of light and the passage
time of a light ray through an inhomogeneous medium.

Our task is to calculate measurable quantities from Eq.\ (\ref{3.2}). 
A very general one is given by the spatial and temporal autocorrelation
function for the electric light field, $\langle \bbox{E}(\bbox{r}_1,t_1) 
\otimes \bbox{E}(\bbox{r}_2,t_2) \rangle$, which we write with the help of
center-of-``mass'' ($\bbox{R},T$) and relative ($\bbox{r},t$) coordinates:
\begin{eqnarray}
\lefteqn{\bbox{W}(\bbox{R},\bbox{r},T,t) \,=\qquad \qquad \qquad } \nonumber\\
\label{3.3}
 & & \qquad 
 \Bigl\langle  
 \bbox{E}\Bigl(\bbox{R}+\frac{\bbox{r}}{2}, T+\frac{t}{2}\Bigr)\otimes \, 
  \bbox{E}^{\ast}\Bigl(\bbox{R}-\frac{\bbox{r}}{2}, T-\frac{t}{2}\Bigr) 
  \Bigr\rangle \enspace.
\end{eqnarray}
It is a second rank tensor. 
For $\bbox{r}=\bbox{0}$ we derive the scalar
\begin{equation}
\label{3.4}
 W(\bbox{R},T,t) = \Bigl\langle 
        \bbox{E}\Bigl(\bbox{R}, T+\frac{t}{2}\Bigr) \cdot \bbox{\varepsilon}_0
     \bbox{E}^{\ast}\Bigl(\bbox{R}, T-\frac{t}{2}\Bigr) \Bigr\rangle \enspace,
\end{equation}
which at $t=0$ is equal to the energy density of the light field when
the small fluctuating part $\delta \bbox{\varepsilon}$ is neglected.
The $T$ dependence
is, e.g., due to time-modulated light sources, and $\bbox{R}$ describes
variations of the energy density on long length scales.
For $t \ne 0$, $W(\bbox{R},T,t)$ reflects the dynamics of the scattering medium
through its dependence on the structure factor $\bbox{B}^{\omega}(\bbox{r},t)$.
It is measured in dynamic 
light scattering experiments either in single scattering \cite{Berne76}
or with DWS \cite{Weitz92}. 
We will see below that the Fourier transform
with respect to $\bbox{r}$, $\bbox{W}(\bbox{R},\bbox{k},T,t)$, is diagonal in 
the basis $\{ \bbox{e}_{\alpha}(\hat{\bbox{k}}) \otimes 
 \bbox{e}_{\beta}(\hat{\bbox{k}}) \}$ on length ($\bbox{R}$) and time ($T$)
scales much longer than the wavelength and time period of light,
\begin{equation}
\label{3.5}
\bbox{W}(\bbox{R},\bbox{k},T,t) \approx \sum_{\alpha}
 W^{\alpha}(\bbox{R},\bbox{k},T,t) \, \bbox{e}_{\alpha}(\hat{\bbox{k}}) 
 \otimes \bbox{e}_{\alpha}(\hat{\bbox{k}}) \enspace.
 \end{equation}
For $t=0$ we can, therefore, interpret $W^{\alpha}(\bbox{R},\bbox{k},T,t)$
as the energy density of a light wave with wave vector $\bbox{k}$ and
polarization $\bbox{e}_{\alpha}(\hat{\bbox{k}})$. To clarify this 
interpretation we look at the Fourier transform
\begin{eqnarray}
\lefteqn{\int d^3\bbox{R} d^3\bbox{r} dT dt \, \bbox{W}(\bbox{R},\bbox{r},T,t)
\,
e^{i(-\bbox{K}\cdot \bbox{R}-\bbox{k}\cdot \bbox{r} + \Omega T + \omega t)}
 } \nonumber \\
\label{3.5a}
 & & = \,\Bigl\langle \bbox{E}\Bigl(\bbox{k}+\frac{\bbox{K}}{2}, \omega + 
 \frac{\Omega}{2}\Bigr) \otimes 
  \bbox{E}^{\ast}\Bigl(\bbox{k}-\frac{\bbox{K}}{2}, \omega-\frac{\Omega}{2}
 \Bigr) \Bigr\rangle \enspace.
\end{eqnarray}
Since $\bbox{R}$ and $T$ describe variations on long scales we have
$K \ll k$ and $\Omega \ll \omega$. Thus the amplitudes 
$\bbox{E}(\bbox{k}\pm \bbox{K}/2, \omega \pm \Omega/2)$ are srongly peaked 
around $\omega$ and $\bbox{k}$ which can be identified, respectively, with 
the frequency and wave vector of the light waves in the medium.

To calculate the autocorrelation function $\bbox{W}(\bbox{R},\bbox{r},T,t)$ 
for special light sources $\bbox{W}_0(\bbox{R},\bbox{r},T,t)$ and/or given 
boundary conditions, we need the averaged ``two-particle'' Green function 
\begin{equation}
\label{3.6}
 \bbox{\Phi} = \langle \, 
   \bbox{G}^{R} \otimes \bbox{G}^{A} \, \rangle^{(N)} \enspace ,
\end{equation}
where $\bbox{G}^{R}$ and $\bbox{G}^{A}$ denote the retarded
and advanced Green functions with $\bbox{G}^{A} = 
[\bbox{G}^{R}]^{\ast}$. Then
\begin{equation}
\bbox{W}(1) = \int d2\, \Phi(1,2) \bbox{W}_0(2)
\end{equation}
(for notation see Appendix \ref{sec app1}).
In the second subsection we will derive
the diffusion pole of $\bbox{\Phi}$, i.e., that part of $\bbox{\Phi}$
which corresponds to a diffusion equation for $\bbox{W}(\bbox{R},\bbox{k},T,t)$
in the variables $\bbox{R}$ and $T$. 
It governs the propagation of light energy at long length and time scales.
The time $t$ will appear in an absorption term of the diffusion equation, 
which is zero for $t=0$. In the first subsection we calculate the averaged 
retarded and advanced one-particle Green functions, which we need for the 
derivation of $\bbox{\Phi}$.

\subsection{The averaged one-particle Green function} 
 \label{subsec rand.oneG}

\begin{figure}
\centerline{\psfig{figure=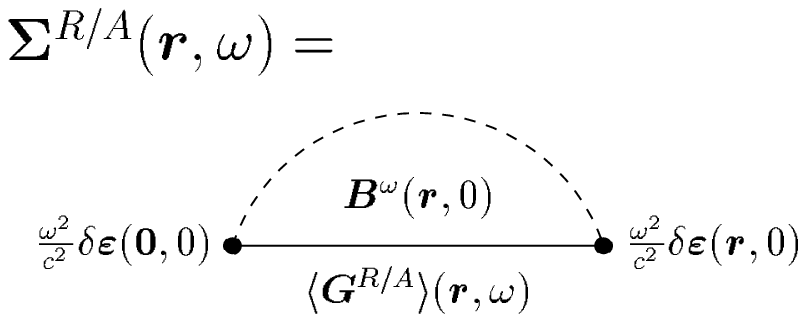}}

\vspace{.5cm}

\caption[]{The mass operators $\bbox{\Sigma}^{R/A}(\bbox{r},\omega)$ in the 
weak-scat\-ter\-ing approximation consist of two
scat\-ter\-ing events from inhomogeneities in the dielectric tensor, which are
tied together by spatial correlations [$\bbox{B}^{\omega}(\bbox{r},t=0)$]
and the averaged propagators $\langle \bbox{G}^{R/A} \rangle (\bbox{r},\omega)$
of the electric field.}
\label{fig2a}
\end{figure}

The one-particle Green functions $\langle \bbox{G}^{R} \rangle$ and
$\langle \bbox{G}^{A} \rangle$ follow from Dyson's equation \cite{Frisch68},
which we give
in a formal notation:\footnote{In the following we will use a coordinate-free 
representation for tensors and their contractions which we explain here. 
If $\bbox{A}$ and $\bbox{B}$ are second rank tensors so is $\bbox{A}\bbox{B}$ 
with components $[\bbox{A}\bbox{B}]_{ij} = \sum_{k}A_{ik}B_{kj}$. For fourth 
rank tensors $\bbox{C}$ and $\bbox{D}$ we form a tensor $\bbox{C}\bbox{D}$ of 
the same rank with components $[\bbox{C}\bbox{D}]_{ijkl} = 
\sum_{mn} C_{ijmn} D_{mnkl}$. Finally, $\bbox{C}\bbox{A}$ is a second rank
tensor and has components $[\bbox{C}\bbox{A}]_{ij}= \sum_{kl} C_{ijkl}A_{kl}$.}
\begin{equation}
\label{3.7}
\langle \bbox{G}^{R/A} \rangle = \bbox{G}_0 + \bbox{G}_0 \, 
\bbox{\Sigma}^{R/A} \, \langle \bbox{G}^{R/A} \rangle \enspace.
\end{equation}
We have introduced the retarded and advanced second-rank-tensor mass operators
$\bbox{\Sigma}^{R}$ and $\bbox{\Sigma}^{A}$.
In the weak-scattering approximation, where the elements of the structure 
factor $\bbox{B}^{\omega}(\bbox{r},t=0)$ are assumed to be much smaller than 
one, these mass operators are proportional to 
$\bbox{B}^{\omega}(\bbox{r},t=0)$ and can be written in frequency space as
\begin{equation}
\label{3.8}
\bbox{\Sigma}^{R/A}(\bbox{r},\omega) = \bbox{B}^{\omega}(\bbox{r},t=0)
\langle \bbox{G}^{R/A} \rangle (\bbox{r},\omega) \enspace .
\end{equation}
Figure\ \ref{fig2a} gives a diagrammatic representation of 
$\bbox{\Sigma}^{R/A}(\bbox{r},\omega)$.
%
%
Finally, in momentum space, we get from Eq.\ (\ref{3.7}),
\begin{equation}
\label{3.9}
\langle \bbox{G}^{R/A} \rangle (\bbox{k},\omega) = 
[\bbox{G}_0^{-1}(\bbox{k},\omega) - \bbox{\Sigma}^{R/A}(\bbox{k},\omega) ]^{-1}
\enspace.
\end{equation}
The inverse Green function $\bbox{G}_0^{-1}(\bbox{k},\omega)$ is diagonal
in the basis $\{\bbox{d}^{\,i}(\hat{\bbox{k}}) \otimes 
\bbox{d}^{\,j}(\hat{\bbox{k}}) \}$, and
$\bbox{\Sigma}^{R/A}(\bbox{k},\omega)$ causes a small pertubation. From 
perturbation theory we know that to zeroth order in $\bbox{\Sigma}^{R/A}$
the eigenvectors of $\bbox{G}_0^{-1}(\bbox{k},\omega) - 
\bbox{\Sigma}^{R/A}(\bbox{k},\omega)$ are unchanged, whereas to first order
in 
$\hspace*{-.3cm}$$\bbox{\Sigma}^{R/A}$ the diagonal elements
\begin{equation}
\label{3.10}
[\bbox{\Sigma}^{R/A}(\bbox{k},\omega)]_{i} =
 \bbox{e}_{i}(\hat{\bbox{k}}) \cdot \bbox{\Sigma}^{R/A}(\bbox{k},\omega)
 \bbox{e}_{i}(\hat{\bbox{k}})
\end{equation}
contribute to the eigenvalues of $\langle \bbox{G}^{R/A} \rangle$ 
renormalizing the wave numbers $\frac{\omega}{c}\, n_i$ of the light modes.
The significant effect of $\bbox{\Sigma}^{R/A}$ comes from its imaginary
part. Then, in its final form the propagating part of the Green functions 
reads
\begin{equation}
\label{3.11}
\langle \bbox{G}^{R/A}\rangle(\bbox{k},\omega) \approx 
\sum_{\alpha=1}^{2} [\langle \bbox{G}^{R/A}\rangle(\bbox{k},\omega)
]^{\alpha} \, \bbox{e}_{\alpha}(\hat{\bbox{k}}) \otimes 
\bbox{e}_{\alpha}(\hat{\bbox{k}})
\end{equation}
with
\begin{equation}
\label{3.11a}
[\langle \bbox{G}^{R/A}\rangle(\bbox{k},\omega)]^{\alpha} =
\Bigl[ \frac{\omega^2}{c^2} - \frac{k^2}{n_{\alpha}^2(\hat{\bbox{k}})} 
\mp \frac{i\omega}{c n_{\alpha}(\hat{\bbox{k}}) 
l_{\alpha}(\hat{\bbox{k}},\omega)} \Bigr]^{-1} .
\end{equation}
We have introduced the scattering mean-free-path 
$l_{\alpha}(\hat{\bbox{k}},\omega)$ of the light mode 
$\{\bbox{k}^{\alpha}\,|\, \bbox{e}_{\alpha}(\hat{\bbox{k}}) \}$ which travels
with a wave vector 
$\bbox{k}^{\alpha} = \frac{\omega}{c}n_{\alpha} \hat{\bbox{k}}$ and
polarization $\bbox{e}_{\alpha}(\hat{\bbox{k}})$:
\begin{equation}
\label{3.12}
l_{\alpha}(\hat{\bbox{k}},\omega) = \Bigl[-\frac{c}{\omega} \, 
 n_{\alpha}(\hat{\bbox{k}})
 [\text{Im} \bbox{\Sigma}^{R}( {\textstyle \frac{\omega}{c}} 
 n_{\alpha}\hat{\bbox{k}},
\omega)]_{\alpha} \Bigr]^{-1} \enspace.
\end{equation}
To arrive at $l_{\alpha}(\hat{\bbox{k}},\omega)$, we used a CPA approximation
replacing the argument $\bbox{k}$ of 
$[\bbox{\Sigma}^{R/A}(\bbox{k},\omega)]_{\alpha}$
by the wave vector $\frac{\omega}{c} \, n_{\alpha}\hat{\bbox{k}}$ of the
light modes where $\bbox{G}_0$ diverges. 
$\text{Im}\bbox{\Sigma}^{R}$ follows from
Eq.\ (\ref{3.8}) using the momentum shell approximation
\begin{equation}
\label{3.13}
[\text{Im} \langle \bbox{G}^{R/A} \rangle (\bbox{k},\omega)]^{\alpha} 
 \approx -\frac{\pi}{2}\frac{c}{\omega} \, n_{\alpha}(\hat{\bbox{k}})
 \delta({\textstyle \frac{\omega}{c}}\, n_{\alpha}(\hat{\bbox{k}}) - k)
\end{equation}
valid in the limit $\bbox{\Sigma}^{R} \rightarrow \bbox{0}$.
We finally obtain
\begin{equation}
\label{3.14}
l_{\alpha}(\hat{\bbox{k}},\omega) =  \biggl[\frac{\pi}{2} \,
n_{\alpha}(\hat{\bbox{k}}) \sum_{\beta=1}^{2} \int_{\hat{\bbox{q}}^{\beta}} 
[\bbox{B}^{\omega}_{\bbox{k}^{\alpha}\bbox{q}^{\beta}}(t=0)]_{\alpha\beta}
 \biggr]^{-1} \enspace,
\end{equation}
where we sum over all possible scattering events of the incoming light mode
$\{ \bbox{k}^{\alpha}\, | \, \bbox{e}_{\alpha}(\hat{\bbox{k}}) \}$ into modes
$\{ \bbox{q}^{\beta} \, | \, \bbox{e}_{\beta}(\hat{\bbox{k}}) \}$. The single
scattering event is described by the structure factor
\begin{equation}
\label{3.15}
[\bbox{B}^{\omega}_{\bbox{k}^{\alpha}\bbox{q}^{\beta}}(t=0)]_{\alpha\beta}
 = \frac{\omega^4}{c^4} \, \langle \, |\bbox{e}_{\alpha}(\hat{\bbox{k}}) \cdot
\delta \bbox{\varepsilon}(\bbox{q}_s,0)
\bbox{e}_{\beta}(\hat{\bbox{q}})|^2 \,\rangle \enspace,
\end{equation}
which is proportional to the differential scattering cross section.
We also introduced the scattering vector
\begin{equation}
\label{3.15a}
\bbox{q}_s = \frac{\omega}{c} \, (n_{\alpha}\hat{\bbox{k}} 
 - n_{\beta}\hat{\bbox{q}}) \enspace.
\end{equation}
In Eq.\ (\ref{3.14}) and throughout the article, we use a shorthand notation
for angular integration:
\begin{equation}
\label{3.16}
\int_{\hat{\bbox{q}}^{\beta}} \ldots \enspace = 
\int \frac{d\Omega_{\bbox{q}}}{(2\pi)^3}
n_{\beta}^3(\hat{\bbox{q}}) \,\, \ldots 
\enspace.
\end{equation}
We point out that the scattering mean-free-path 
$l_{\alpha}(\hat{\bbox{k}},\omega)$ now depends on the direction 
$\hat{\bbox{k}}$ and the polarization $\bbox{e}_{\alpha}(\hat{\bbox{k}})$ of 
the light mode.
Following the work of Nelson and Lax \cite{Lax73}, it is straightforward
to show that $l_{\alpha}(\hat{\bbox{k}},\omega)$ determines an exponential
decay of the far field of $\langle \bbox{G}^{R}\rangle(\bbox{R},\omega)$
travelling in the direction $\hat{\bbox{k}}$ with polarization 
$\bbox{e}_{\alpha}(\hat{\bbox{k}})$.
In our analysis, we have neglected the off-diagonal components of 
$\bbox{\Sigma}^{R/A}$, so that the polarization vectors of our light rays
are identical to those of the homogeneous medium.
We will use this approximation in the next subsection.
We conclude with the definition of the two quantities
\begin{equation}
\label{3.16a}
\Delta\bbox{G}_{\bbox{k}}^{\omega}(\bbox{K},\Omega) =  
\langle \bbox{G}^{R}\rangle (\bbox{k}_{+}, \omega_{+})
- \langle \bbox{G}^{A}\rangle(\bbox{k}_{-}, \omega_{-}) \enspace,
\end{equation}
and
\begin{equation}
\label{3.16b}
\Delta\bbox{\Sigma}_{\bbox{k}}^{\omega}(\bbox{K},\Omega) =  
\bbox{\Sigma}^{R} (\bbox{k}_{+}, \omega_{+})
-  \bbox{\Sigma}^{A} (\bbox{k}_{-}, \omega_{-}) \enspace,
\end{equation}
where
\begin{equation}
\label{3.16c}
\bbox{k}_{\pm}=\bbox{k}\pm \bbox{K}/2 \quad \text{and} 
\quad \omega_{\pm}=\omega \pm \Omega/2 \enspace.
\end{equation}
We will use both of these in the following, especially when 
$\bbox{K}=\bbox{0}$ and $\Omega=0$:
\begin{equation}
\label{3.16d}
\Delta\bbox{G}_{\bbox{k}}^{\omega}(\bbox{0},0) =
2 i \text{Im} \langle\bbox{G}^{R}\rangle (\bbox{k},\omega)
\end{equation}
and
\begin{equation}
\label{3.16e}
\Delta\bbox{\Sigma}_{\bbox{k}}^{\omega}(\bbox{0},0)=
2 i \text{Im} \bbox{\Sigma}^{R} (\bbox{k},\omega) 
=-\frac{\omega}{c} \, 
\frac{2i}{n_{\alpha}(\hat{\bbox{k}}) l_{\alpha}(\hat{\bbox{k}},\omega)}
\enspace.
\end{equation}

\subsection{The averaged two-particle Green function} \label{subsec rand.twoG}

\begin{figure}
\centerline{\psfig{figure=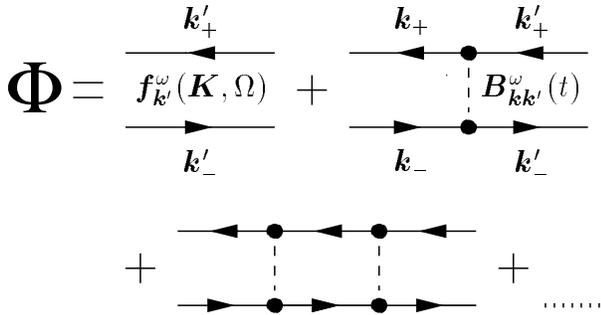}}

\vspace*{.5cm}

\caption[]{The two-particle Green function $\bbox{\Phi}$ as a sum of ladder
   diagrams. $\bbox{f}^{\omega}_{\bbox{k}}(\bbox{K},\Omega)$ propagates
   two electric field modes with wave vectors $\bbox{k}_{+}'$ and 
   $\bbox{k}_{-}'$. Both are scattered by the same inhomogeneity in the
   dielectric tensor described by the structure factor 
   $\bbox{B}^{\omega}_{\bbox{k}\bbox{k}'}(t)$ for the scattering vector 
   $\bbox{k} - \bbox{k}'$.}
\label{fig2}
\end{figure}

The averaged two-particle Green function $\bbox{\Phi}$ obeys the 
Bethe-Salpeter equation \cite{Frisch68}:
\begin{equation}
\label{3.17}
\bbox{\Phi} = [\langle \bbox{G}^{R} \rangle \otimes  \langle \bbox{G}^{A} \rangle ]^{(N)}
 + [\langle \bbox{G}^{R} \rangle \otimes  \langle \bbox{G}^{A} \rangle ]^{(N)} \bbox{U}
\bbox{\Phi} \enspace ,
\end{equation}
where $\bbox{U}$ stands for the irreducible vertex function or intensity 
operator, which
in the weak-scattering approximation, equals  the structure factor: 
$\bbox{U} \approx \bbox{B}$. The 
Bethe-Salpeter equation is best handled in momentum and frequency space.
With all arguments, the Green function is 
$\bbox{\Phi}^{\omega}_{\bbox{k}\bbox{k}'}(\bbox{K},\Omega,t)$ 
(see Appendix \ref{sec app1}).
The variables $\bbox{K}$, $\Omega$ correspond to the center of ``mass''
coordinates $\bbox{R}$, $T$, introduced in the introduction
to this section,  and the wave vectors $\bbox{k}$, $\bbox{k}'$ to
relative coordinates $\bbox{r}$, $\bbox{r}'$.
The superscript $\omega$
is the light frequency, and the $t$ dependence explicitly comes from the
structure factor $\bbox{B}^{\omega}_{\bbox{k}\bbox{k}'}(t)$.
The explicit form of the Bethe-Salpeter equation reads
\begin{eqnarray}
\lefteqn{ \displaystyle
 \int \frac{d^3k_1}{(2\pi)^3} [\bbox{1}_{\bbox{k}\bbox{k}_1}^{(4)} - 
\bbox{f}^{\omega}_{\bbox{k}}(\bbox{K},\Omega)\,
\bbox{B}^{\omega}_{\bbox{k}\bbox{k}_1}(t)] 
\bbox{\Phi}^{\omega}_{\bbox{k}_1\bbox{k}'}(\bbox{K},\Omega,t)} \nonumber\\
\label{3.18}
& &\hspace*{4cm} = \, \bbox{f}^{\omega}_{\bbox{k}}(\bbox{K},\Omega) 
\bbox{1}_{\bbox{k}\bbox{k}'}^{(4)} \enspace,
\end{eqnarray}
which we derive in Appendix \ref{sec app1}.  In the last equation  we have 
introduced
an abbreviation for the tensor product of the averaged one-particle
Green functions:
\begin{equation}
\label{3.19}
\bbox{f}^{\omega}_{\bbox{k}}(\bbox{K},\Omega) =  \bigl[\langle
\bbox{G}^{R}\rangle (\bbox{k}_{+} ,\omega_{+} )  \otimes\, 
\langle \bbox{G}^{A} \rangle (\bbox{k}_{-},\omega_{-}) \bigr]^{(N)}  \enspace,
\end{equation}
and a shorthand notation $\bbox{1}_{\bbox{k}\bbox{k}'}^{(4)}$ for the 
combination of the delta function and the unit element of fourth rank tensors:
\begin{equation}
\label{3.20}
[\bbox{1}_{\bbox{k}\bbox{k}'}^{(4)}]_{ijkl} := (2\pi)^3 \,
\delta(\bbox{k}-\bbox{k}')\,(\delta_{ik}\delta_{jl}+\delta_{il}
\delta_{jk}) / 2 \enspace.
\end{equation}
Here $i,j,k$ and $l$ are Cartesian indices. The Bethe-Salpeter equation 
(\ref{3.18}) can be solved formally by iteration, leading to a sum of ladder
diagrams  (see Fig.\ \ref{fig2}).
$\hspace*{-.4cm}$The multiple integrals and the sum can be
done analytically for $\delta$-function correlations for 
$\bbox{K} \rightarrow \bbox{0}$, and the solution represents the diffusion 
pole of $\bbox{\Phi}$. For the anisotropic, long-range correlations of our 
problem this procedure is not possible. Vollhardt and W\"olfle
\cite{Vollhardt80} derived the diffusion pole for isotropic electron 
transport directly from Eq.\ (\ref{3.18}), and MacKintosh and John 
\cite{MacKintosh89} applied their method to light. 
In the anisotropic case one has to be more careful, as W\"olfle and Bhatt
showed for electrons \cite{Woelfle84}. We will proceed a similar way, however 
our problem is more difficult because we have to deal with the different
polarizations of light. 

Let $\bbox{\Psi}_{\bbox{k}}^{(n)}(\bbox{K},\Omega,t)$
and $\lambda^{(n)} (\bbox{K},\Omega,t)$ be, respectively, the $n$th 
eigenvector and eigenvalue of the integral operator of Eq.\ (\ref{3.18}),
\begin{equation}
\label{3.21}
 \int \frac{d^3k_1}{(2\pi)^3} [\bbox{1}_{\bbox{k}\bbox{k}_1}^{(4)} - 
\bbox{f}^{\omega}_{\bbox{k}}(\bbox{K},\Omega)\,
\bbox{B}^{\omega}_{\bbox{k}\bbox{k}_1}(t)]\bbox{\Psi}_{\bbox{k}_1}^{(n)} 
= \lambda^{(n)} \bbox{\Psi}_{\bbox{k}}^{(n)} \enspace,
\end{equation}
and $\overline{\bbox{\Psi}}_{\bbox{k}}^{(n)}(\bbox{K},\Omega,t)$ the 
eigenvectors of the Hermitian adjoint operator \cite{Morse53}. Then it is 
straightforward to show \cite{MacKintosh89}, with the aid of the completeness
relation
\begin{equation}
\label{3.23}
\sum_{n} \bbox{\Psi}_{\bbox{k}}^{(n)} \otimes 
\overline{\bbox{\Psi}}_{\bbox{k}'}^{(n)} = 
\bbox{1}_{\bbox{k}\bbox{k}'}^{(4)} \enspace,
\end{equation}
that
\begin{equation}
\label{3.22}
\bbox{\Phi}^{\omega}_{\bbox{k}\bbox{k}'}(\bbox{K},\Omega,t) = \sum_{n}
\frac{\bbox{\Psi}_{\bbox{k}}^{(n)} \otimes 
\overline{\bbox{\Psi}}_{\bbox{k}'}^{(n)}}{\lambda^{(n)} 
} \, \bbox{f}^{\omega}_{\bbox{k}'}(\bbox{K},\Omega)
\end{equation}
solves the Bethe-Salpeter equation.
When $\bbox{K}=\bbox{0}$ and $\Omega=t=0$, the quantity 
$\Delta\bbox{G}_{\bbox{k}}^{\omega}(\bbox{0},0)$, defined in 
Eq.\ (\ref{3.16a}), is an eigenvector of Eq.\ (\ref{3.21}) with eigenvalue
$\lambda^{(0)}(\bbox{0},0,0) = 0$. To prove this statement we use a special
case of one of the Ward identities (see Appendix \ref{subsec app2.ward}),
\begin{equation}
\label{3.24}
\Delta \bbox{\Sigma}_{\bbox{k}}^{\omega}(\bbox{0} , 0 ) =
\int \frac{d^3 k'}{(2\pi)^3} \, \bbox{B}_{\bbox{k}\bbox{k}' }^{\omega}(t=0) \,
 \Delta \bbox{G}_{\bbox{k}'}^{\omega}(\bbox{0} , 0 )  \enspace,
\end{equation}
and the relation
\begin{equation}
\label{3.25}
\Delta \bbox{G}_{\bbox{k}}^{\omega}(\bbox{0} , 0) =
\bbox{f}_{\bbox{k}}^{\omega}(\bbox{0} , 0) \, 
\Delta \bbox{\Sigma}_{\bbox{k}}^{\omega}(\bbox{0} , 0) \enspace.
\end{equation}
The first equation is obvious from the definition of the mass operator, and
the second one is given in Appendix \ref{subsec app2.use}.
We have identified the diffusion pole, as we shall explicitly see soon.
The result is valid beyond the weak-scattering approximation and based on the
Ward identities \cite{Vollhardt80}. All other eigenvalues of Eq.\ (\ref{3.21})
are positive, and in the real-space coordinate $\bbox{R}$, they give 
exponentially decaying contributions to 
$\bbox{\Phi}^{\omega}_{\bbox{k}\bbox{k}'}(\bbox{K},\Omega,t)$, which are not 
important at long length scales \cite{MacKintosh89}. 

The diffusion
approximation follows when we calculate $\lambda^{(0)}(\bbox{K},\Omega,t)$
with the help of perturbation theory in the limit $K,\Omega,t \rightarrow 0$.
First, we have to get the eigenvalue equation in this limit.
To proceed we need the following equation, which we derive in 
Appendix \ref{subsec app2.use}:
\begin{eqnarray}
[\Delta \bbox{G}_{\bbox{k}}^{\omega}(\bbox{K} , \Omega)]^{\alpha} & \approx &
[\bbox{f}_{\bbox{k}}^{\omega}(\bbox{0} , 0)]^{\alpha \alpha} \, 
\Bigl( [\Delta \bbox{\Sigma}_{\bbox{k}}^{\omega}(\bbox{0} ,  0)]_{\alpha} 
 \nonumber \\
\label{3.26}
 & & \qquad
 - \frac{\partial [\bbox{G}_0^{-1}]_{\alpha} }{\partial \bbox{k} } \cdot 
 \bbox{K} - \frac{2\omega}{c^2} \, \Omega \Bigr) \enspace .
\end{eqnarray}
As usual, the Greek superscript or subscript $\alpha$ refers, respectively, to 
the basis "vectors" $\bbox{e}_{\alpha}(\hat{\bbox{k}}) \otimes 
\bbox{e}_{\alpha}(\hat{\bbox{k}})$ or $\bbox{d}^{\alpha}(\hat{\bbox{k}}) 
\otimes \bbox{d}^{\alpha}(\hat{\bbox{k}})$. Equation\ (\ref{3.26}) gives 
$[\Delta \bbox{G}_{\bbox{k}}^{\omega}(\bbox{K} , \Omega)]^{\alpha}$ to first
order in $\bbox{K}$, $\Omega$ and $\bbox{\Sigma}$. It shows clearly where
the wave vector $\bbox{K}$ and the frequency $\Omega$ come in. Now, we
formulate eigenvalue equation (\ref{3.21}) in components, choosing 
$\bbox{K} = \bbox{0}$ and $\Omega=0$. Then, we apply the last equation and 
arrive at
\begin{eqnarray}
 &\displaystyle \Bigl( 
 [\Delta \bbox{\Sigma}_{\bbox{k}}^{\omega}(\bbox{0},0)]_{\alpha} 
-\frac{\partial [\bbox{G}_0^{-1}]_{\alpha} }{\partial \bbox{k} } \cdot 
 \bbox{K} - \frac{2\omega}{c^2} \, \Omega \Bigr) \,
\Psi_{\bbox{k}}^{\alpha} & \nonumber \\
\label{3.27}
 &\displaystyle -\,[\Delta \bbox{G}_{\bbox{k}}^{\omega}(\bbox{0} , 0)]^{\alpha}
 \sum_{\beta} \int \frac{d^3k_1}{(2\pi)^3} 
[\bbox{B}^{\omega}_{\bbox{k}\bbox{k}_1}(t)]_{\alpha \beta}  
\Psi_{\bbox{k}_1}^{\beta} & \\
 & \displaystyle = \, \lambda(\bbox{K},\Omega,t) 
 [\Delta \bbox{\Sigma}_{\bbox{k}}^{\omega}(\bbox{0},0)]_{\alpha} 
 \Psi_{\bbox{k}}^{\alpha} & \nonumber
\end{eqnarray}
as the eigenvalue equation, which is correct up to first order in
$\bbox{K}$, $\Omega$, and $\bbox{\Sigma}$.
In a next step we use the ansatz
\begin{equation}
\label{3.28}
\Psi^{\alpha}_{\bbox{k}} \propto
[\Delta \bbox{G}_{\bbox{k}}^{\omega}(\bbox{0},0)]^{\alpha} \,
 \bigl[\widetilde{\Psi}_0 \pi + 
\sum_{i=1}^{\ldots}
\widetilde{\Psi}_{i} \, \varphi_{i}^{\alpha}(\hat{\bbox{k}})
\bigr]
\enspace,
\end{equation}
for the eigenfunctions to turn the eigenvalue equation into a matrix equation.
The first factor on the right-hand side forces $k$ to equal 
$\frac{\omega}{c} n_{\alpha}(\hat{\bbox{k}})$ in the momentum shell
approximation:
\begin{equation}
\label{3.29}
[\Delta\bbox{G}_{\bbox{k}}^{\omega}(\bbox{0},0)]^{\alpha} \approx
 -i\pi\frac{c}{\omega} n_{\alpha}(\hat{\bbox{k}}) \,
 \delta( \textstyle{\frac{\omega}{c}} n_{\alpha}(\hat{\bbox{k}})-k ) \enspace.
\end{equation}
The amplitude $\widetilde{\Psi}_0$ represents the zeroth eigenvector, and the
second term involves a complete set of real basis functions
$\bbox{\varphi}_{i}(\hat{\bbox{k}})$ on the unit sphere and for the space 
spanned by $\bbox{e}_1 \otimes \bbox{e}_1$ and $\bbox{e}_2 \otimes \bbox{e}_2$.
We will comment below on how to choose these basis functions.
Keeping only dominant terms in the small $\Omega$, $K$, and $t$ limit, we can
write the eigenvalue equation in matrix form as
\begin{eqnarray}
&\left[ 
 \begin{array}{cc}
    \frac{2\overline{n^3}}{\pi c\rule[-2mm]{0mm}{2mm}}\,
    [-i\Omega+\mu(\omega,t)] & 
    -i[{\cal G}(\bbox{K})]^{\text{t}} \\
    -i {\cal G}(\bbox{K}) & {\cal B}
 \end{array}
 \right] 
     \, \left[
        \begin{array}{c}
           \widetilde{\Psi}_0 \\
           \vdots
        \end{array}
        \right]\rule[-4mm]{0mm}{4mm} \qquad \enspace &  \nonumber\\
\label{3.30}
 & \quad \qquad \qquad = \, \lambda(\bbox{K},\Omega,t) \, 
 \left[
 \begin{array}{cc}
   \Sigma & {\cal S}_{0}^{\text{t}} \\
   {\cal S}_{0} & {\cal S}_{1}
 \end{array}
 \right]
     \, \left[
        \begin{array}{c}
           \widetilde{\Psi}_0 \\
           \vdots
        \end{array}
        \right] \enspace .  &
\end{eqnarray}
The matrix ${\cal B}$ basically represents the structure factor 
$[\bbox{B}^{\omega}_{\bbox{k}^{\alpha}\bbox{q}^{\beta}}(0)]_{\alpha \beta}$ in
our chosen basis,
\begin{eqnarray}
\protect[\,{\cal B}\,]_{ij} & = & 
\pi \, \sum_{\alpha,\beta} \Bigl\{
 \int_{\hat{\bbox{k}}^{\alpha}}\int_{\hat{\bbox{q}}^{\beta}} 
\bigl[ \varphi_{i}^{\alpha}(\hat{\bbox{k}}) 
 \varphi_{j}^{\alpha}(\hat{\bbox{k}}) \nonumber \\
\label{3.31}
& &
- \varphi_{i}^{\alpha}(\hat{\bbox{k}})
 \varphi_{j}^{\beta}(\hat{\bbox{q}}) \bigr] \, 
[\bbox{B}^{\omega}_{\bbox{k}^{\alpha}\bbox{q}^{\beta}}(0)]_{\alpha \beta}
\Bigr\}\enspace.
\end{eqnarray}
The vector ${\cal G}$ with components
\begin{equation}
\label{3.32}
 \protect[{\cal G}(\bbox{K})]_{i} = \pi \, \sum_{\alpha} 
\int_{\hat{\bbox{k}}^{\alpha}} n_{\alpha}(\hat{\bbox{k}})  
\left. \frac{\partial [\bbox{G}_0^{-1}]_{\alpha}}{\partial \bbox{k}}
\right|_{\hat{\bbox{k}}} \cdot \bbox{K} \,
\varphi_{i}^{\alpha}(\hat{\bbox{k}})
\end{equation}
is linear in $\bbox{K}$.
Another important term is the coefficient
\begin{equation}
\label{3.33}
\mu(\omega,t) = \frac{c\pi^3}{2\overline{n^3}}
 \, \sum_{\alpha,\beta} 
\int_{\hat{\bbox{k}}^{\alpha}}\int_{\hat{\bbox{q}}^{\beta}} 
[ \bbox{B}^{\omega}_{\bbox{k}^{\alpha}\bbox{q}^{\beta}}(0)
-\bbox{B}^{\omega}_{\bbox{k}^{\alpha}\bbox{q}^{\beta}}(t)
 ]_{\alpha\beta} \enspace,
\end{equation}
where 
\begin{equation}
\label{3.34}
\overline{n^3} = \frac{1}{8\pi}\, \sum_{\alpha} \int d\Omega_{\bbox{k}} \, 
n^3_{\alpha}(\hat{\bbox{k}}) \enspace.
\end{equation}
For reasons that will become clear shortly, we will refer to $\mu(\omega,t)$
as a {\em dynamic absorption coefficient\/}. 
It is an average over dynamical modes
that decay in time. As a result, it increases from zero with increasing 
$t>0$. Finally we have the constant
\begin{eqnarray}
\label{3.35}
\Sigma & = & \pi^3 \sum_{\alpha,\beta} 
\int_{\hat{\bbox{k}}^{\alpha}} \int_{\hat{\bbox{q}}^{\beta}} 
[ \bbox{B}^{\omega}_{\bbox{k}^{\alpha}\bbox{q}^{\beta}}(0)]_{\alpha\beta} \\
 & = & -\pi \frac{c^2}{\omega^2} \sum_{\alpha} \int \frac{d^3k}{(2\pi)^3} \, 
  \nonumber \\
 & & \qquad \qquad \quad 
 [\Delta \bbox{\Sigma}_{\bbox{k}}^{\omega}(\bbox{0} ,  0)]_{\alpha} 
  [\Delta\bbox{G}_{\bbox{k}}^{\omega}(\bbox{0},0)]^{\alpha} \,\, .
\end{eqnarray}
The second equation above states that $\Sigma$ is proportional to the 
normalization
factor of the eigenfunction $\Delta\bbox{G}_{\bbox{k}}^{\omega}(\bbox{0},0)$,
with $\Delta \bbox{\Sigma}_{\bbox{k}}^{\omega}(\bbox{0},0)$ being the 
eigenfunction of the Hermitian conjugated problem. The vector 
${\cal S}_{0}$ and the matrix ${\cal S}_{1}$ are irrelevant here.
For $K,\Omega,t=0$ the zeroth eigenvalue $\lambda^{(0)}$ corresponds to the 
eigenvector $[1,0,0,\ldots]$ in the eigenvalue equation\ (\ref{3.30}), and we
want to know how it evolves for small $K$, $\Omega$ and $t$. The vector
${\cal G}(\bbox{K})$ couples the 00-element with the matrix ${\cal B}$. The
coupling can be removed by an orthogonal transformation with
\begin{equation}
\label{3.36}
{\cal U} = \left[
             \begin{array}{cc}
               1 & i[{\cal G}(\bbox{K})]^{\text{t}} {\cal B}^{-1}\\
               -i {\cal B}^{-1} {\cal G}(\bbox{K}) & \bbox{1}
             \end{array}
           \right] \enspace,
\end{equation}
which essentially renormalizes the 00-element. Calculating $\lambda^{(0)}$ 
from the transformed system to leading order in $K^2$ shows that
$\lambda^{(0)}\Sigma$ is equal to the renormalized 00-element, or
\begin{equation}
\label{3.37}
\lambda^{(0)}
= \frac{2\overline{n^3}}{c \Sigma}\,
\Bigl[-i\Omega + \mu(\omega,t) + \frac{\pi c}{2\overline{n^3}}\,
{\cal G}(\bbox{K}) \cdot {\cal B}^{-1}{\cal G}(\bbox{K}) \Bigr] \enspace.
\end{equation}
Now, we are ready to write down the diffusion approximation of the averaged 
two-particle Green function from Eq.\ (\ref{3.22}),
\begin{equation}
\bbox{\Phi}^{\omega}_{\bbox{k}\bbox{k}'}(\bbox{K},\Omega,t) \approx
\label{3.38}
\frac{1}{N}
\frac{\Delta\bbox{G}^{\omega}_{\bbox{k}}(\bbox{0},0) \otimes
 \Delta\bbox{G}^{\omega}_{\bbox{k}'}(\bbox{0},0)}{-i\Omega + 
\mu(\omega,t)+\bbox{K}\cdot \bbox{D}(\omega)\bbox{K}}
\end{equation}
with
\begin{equation}
\label{3.39}
\bbox{K} \cdot \bbox{D}(\omega) \bbox{K} =  \frac{c}{2\overline{n^3}} \,
{\cal G}(\bbox{K}) \cdot {\cal B}^{-1} {\cal G}(\bbox{K})
\end{equation}
and $N=-2\overline{n^3}\omega^2/(\pi c^3)$. When $\mu(\omega,t) = 0$ the 
Green function $\bbox{\Phi}^{\omega}_{\bbox{k}\bbox{k}'}(\bbox{K},\Omega,t)$ 
possesses a simple diffusion pole with an anisotropic diffusion tensor
$\bbox{D}(\omega)$. In addition we have
the dynamic absorption coefficient $\mu(\omega,t)$, which forms the basis of
DWS. We will discuss it in the next subsection.

From Eq.\ (\ref{3.39}), it is clear that the problem of calculating 
$\bbox{D}(\omega)$ reduces to the problems of calculating ${\cal G}(\bbox{K})$
and the inverse matrix ${\cal B}^{-1}$. If we had at our disposal
functions $\bbox{\varphi}_{i}(\hat{\bbox{k}})$, that diagonalize ${\cal B}$,
our task would be simple:
\begin{equation}
\label{3.40}
\bbox{K} \cdot \bbox{D}(\omega) \bbox{K} =  \frac{c}{2\overline{n^3}} \,
\sum_i \frac{[{\cal G}(\bbox{K})]_i^2}{[{\cal B}]_{ii}} \enspace .
\end{equation}
Unfortunately these functions are difficult to find in anisotropic systems,
and we are reduced to seeking usable approximation schemes. One scheme
is to choose functions $\bbox{\varphi}_{i}(\hat{\bbox{k}})$ so that only
a few components of ${\cal G}(\bbox{K})$, say $[{\cal G}(\bbox{K})]_i$ for
$i=1,2$, are non-zero. Then, if the off-diagonal components of
$[{\cal B}]_{ij}$, coupling $j$ for $j>2$ to $i$ for $i=1,2$, are small,
$[{\cal G}(\bbox{K})]_i [{\cal B}^{-1}]_{ij} [{\cal G}(\bbox{K})]_j$ can be
approximated by using only the $i,j=1,2$ submatrix of $[{\cal B}]_{ij}$
to calculate $[{\cal B}^{-1}]_{ij}$. We will use this approach in numerical
calculations of $\bbox{D}(\omega)$ in nematic liquid crystals in the next
section.

To test our theory we apply it to isotropic systems. We concentrate on the
case, where fluctuations in the dielectric tensor originate from density 
fluctuations only:
$\delta \bbox{\varepsilon} \propto \delta \varrho \bbox{1}$.
Then the components of the structure factor assume the form 
\begin{equation}
[\bbox{B}^{\omega}_{\bbox{k}^{\alpha}\bbox{q}^{\beta}}(0)]_{\alpha \beta}
= [\bbox{e}_{\alpha}(\hat{\bbox{k}}) \cdot \bbox{e}_{\beta}(\hat{\bbox{q}})]^2
\, B^{\omega}(\text{cos}\vartheta_s) \enspace,
\end{equation}
where $B^{\omega}(\text{cos}\vartheta_s)$ solely depends on the scattering
angle $\vartheta_s$.
In isotropic systems, all the polarization vectors perpendicular to 
$\hat{\bbox{k}}$ are equivalent, and the diffusion constant should not depend 
on the special choice we make. Therefore, to treat the factor 
$[\bbox{e}_{\alpha}(\hat{\bbox{k}}) \cdot \bbox{e}_{\beta}(\hat{\bbox{q}})]^2$,
we can add integrations $\int d\phi_1/(2\pi) \int d\phi_2/(2\pi)$ in the 
formula for $\cal B$.
The angles $\phi_1$ and $\phi_2$, respectively, describe rotations 
of the polarization vectors about $\hat{\bbox{k}}$ and $\hat{\bbox{q}}$.
The integration over $\phi_1$ and $\phi_2$ for a fixed scattering angle 
$\vartheta_s$ is straightforward, and we obtain
\begin{eqnarray}
\lefteqn{\int \frac{d\phi_1}{2\pi} \int \frac{d\phi_2}{2\pi} \,
[\bbox{e}_{\alpha}(\hat{\bbox{k}},\phi_1) \cdot 
\bbox{e}_{\beta}(\hat{\bbox{q}},\phi_2)]^2} 
\nonumber \\
& & \hspace*{4.5cm}
 = \,\frac{1+\text{cos}^2\vartheta_s}{4\varepsilon_0^2} \enspace ,
\end{eqnarray}
where the dielectric constant $\varepsilon_0$ comes in through the
normalization of the polarization vectors.
With this trick we are able to show that ${\cal B}$ is essentially diagonalized
by spherical harmonics\footnote{Our formulas are written for real
  basis functions, a 
generalization to complex ones is straightforward.}.
As basis functions we choose $\bbox{\varphi}_{1lm}(\hat{\bbox{k}})= 
[Y_{lm}(\vartheta,\varphi),Y_{lm}(\vartheta,\varphi)]$ and 
$\bbox{\varphi}_{2lm}(\hat{\bbox{k}})= 
[Y_{lm}(\vartheta,\varphi),-Y_{lm}(\vartheta,\varphi)]$ with 
$\text{cos}\vartheta = \hat{\bbox{k}} \cdot \widehat{\bbox{K}}$. We also
use the addition theorem, $P_l(\text{cos}\vartheta_s) \propto \sum_m 
Y_{lm}^{\ast}(\vartheta,\varphi) Y_{lm}(\vartheta',\varphi')$ \cite{Jackson75},
and arrive at $[{\cal B}]^{\gamma lm}_{\gamma' l' m'} \propto
\delta_{\gamma \gamma'} \delta_{ll'} \delta_{mm'}$. The only non-zero element
of ${\cal G}$ is $[{\cal G}]_{110} = -2\varepsilon_0 K/\sqrt{3}$, and we 
finally find the formula for the diffusion constant:
\begin{eqnarray}
\lefteqn{D \, = \, \frac{16}{3}\,\pi c_0} \nonumber \\
\label{3.41}
 & & \Bigl[ \int d\text{cos}\vartheta_s 
(1- \text{cos}\vartheta_s) (1+ \text{cos}^2\vartheta_s)
B^{\omega}(\text{cos}\vartheta_s) \Bigr]^{-1} \enspace ,
\end{eqnarray}
where $c_0$ stands for the speed of light in the system. Compared to the
scalar case we have an additional factor $1+ \text{cos}^2\vartheta_s$.
$B^{\omega}(\text{cos}\vartheta_s)$ alone describes a
scattering process with incoming and outgoing polarization perpendicular
to the scattering plane (VV scattering), whereas 
$\text{cos}^2\vartheta_s B^{\omega}(\text{cos}\vartheta_s)$ belongs to
scattering with polarizations in the plane (HH scattering) \cite{Berne76}.
In a group theoretical language
$[\bbox{B}^{\omega}_{\bbox{k}^{\alpha}\bbox{q}^{\beta}}(0)]_{\alpha \beta}$
transforms under a high-dimensional identity representation of $SO(3)$.
The relation $[{\cal B}]^{\gamma lm}_{\gamma' l' m'} \propto
 \delta_{ll'} \delta_{mm'}$ then means that we have decomposed 
$[{\cal B}]^{\gamma lm}_{\gamma' l' m'}$ via the irreducible representations 
of $SO(3)$.
For less symmetric phases, like a nematic liquid crystal, we can at least
partially diagonalize $\cal B$ with the help of group theory. The relevant
symmetry group $D_{\infty h}$ has only one-dimensional
representions induced by the functions $\text{exp}(im\varphi)$, where 
$\varphi$ is the azimuthal angle around the symmetry axis. Decomposing 
${\cal B}$ with the help of spherical harmonics gives 
$[{\cal B}]^{\gamma lm}_{\gamma' l' m'} \propto \delta_{mm'}$, where different
$l$ and $l'$ now couple to each other (see next section).

The numerator of the Green function in Eq.\ (\ref{3.38}) contains an
interesting effect. The second factor 
$\Delta\bbox{G}^{\omega}_{\bbox{k}'}(\bbox{0},0)$ collects the light sources
or the incoming light waves. The first factor determines the energy density
$W^{\alpha}(\bbox{R},\bbox{k},T,0)$
of an outgoing light wave independent of the sources. This means that in the
diffusion approximation the outgoing light looses all its correlations with 
the light sources.
Integrating $[\Delta\bbox{G}^{\omega}_{\bbox{k}'}(\bbox{0},0)]_{\alpha}$ over
the wavenumber $k$ shows that the ratio for the energy densities
$W^{\alpha}(\bbox{R},\hat{\bbox{k}},T,0)$ in 
polarization states 1 and 2 is $[n_1(\hat{\bbox{k}})/n_2(\hat{\bbox{k}})]^3$.
Experiments measure light intensities. In Eq.\ (\ref{2.7c}) we learned that
only the projection of the Poynting vector $\bbox{S}^{\alpha}$ on
$\hat{\bbox{k}}$ is simple:
$\bbox{S}^{\alpha} \cdot \hat{\bbox{k}} = (c/n_{\alpha}) 
 W^{\alpha}(\bbox{R},\hat{\bbox{k}},T,0)$. Its ratio for the output
polarizations 1 and 2 is $[n_1(\hat{\bbox{k}})/n_2(\hat{\bbox{k}})]^2$ which
gives the ratio of the output intensities when the Poynting vectors are
parallel to $\hat{\bbox{k}}$. This is the case for light travelling along the
principal axes of the dielectric tensor. The Green function in 
Eq.\ (\ref{3.38}) suggests that there is a diffusion equation for each
light wave with direction $\hat{\bbox{k}}$ and polarization $\alpha$. This
does not mean that we have additional conserved quantities besides the energy
density. It only means that after randomizing the incoming light the 
distribution of the light modes in the light field stays the same.

\subsection{Diffusing Wave Spectroscopy} \label{subsec rand.dws}

If we sum over the two polarization states and integrate over all wave vectors
$\bbox{k}$ the Green function in Eq.\ (\ref{3.38}) is equivalent to a
diffusion equation for the scalar time-correlation function $W(\bbox{R},T,t)$, 
introduced in Eq.\ (\ref{3.4}),
\begin{equation}
\label{3.42}
\Bigl[ \frac{\partial}{\partial T} \, - \bbox{\nabla} \cdot 
\bbox{D}(\omega) \bbox{\nabla} + \mu(\omega,t) \Bigl] \, W(\bbox{R},T,t) =
\varrho(\bbox{R},T) \enspace.
\end{equation}
This equation is the basis of Diffusing Wave Spectroscopy (DWS)
\cite{Stephen88}. Solving it for special boundary
conditions and sources $\varrho(\bbox{R},T)$, which depend on experimental
arrangements, gives $W(\bbox{R},T,t)$ in terms of the 
dynamic absorption coefficient
$\mu(\omega,t)$. In our derivation of the diffusion pole we had to restrict 
ourselves to times $t$ with 
$[ \bbox{B}^{\omega}_{\bbox{k}^{\alpha}\bbox{q}^{\beta}}(0)
-\bbox{B}^{\omega}_{\bbox{k}^{\alpha}\bbox{q}^{\beta}}(t)
 ]_{\alpha\beta} \ll [ \bbox{B}^{\omega}_{\bbox{k}^{\alpha}\bbox{q}^{\beta}}
 (0)]_{\alpha\beta}$. In this time range we expect being able to perform a 
Taylor expansion in $t$ which gives a linear time dependence
for $\mu(\omega,t)$:
\begin{equation}
\label{3.43}
\mu(\omega,t) = \mu_0 t \enspace.
\end{equation}
The constant $\mu_0$ reflects some averaged dynamical properties of the system.
For the diffusion of particles in colloidal suspensions, 
$\mu_0 = 2D_{\text{B}}\omega^2/(l^{\ast}c)$, where $D_{\text{B}}$ is the
self-diffusion constant of the particles \cite{Weitz92}.
The condition just imposed on $t$ means that DWS probes the dynamics of a 
system on a much shorter time scale than single light scattering does, which
requires times where $[ \bbox{B}^{\omega}_{\bbox{k}^{\alpha}\bbox{q}^{\beta}}
 (t)]_{\alpha\beta}$ has already decayed considerably. DWS, therefore, offers
the possibility of studying the validity of the dynamical description of the
system on short time scales. This was done for colloidal suspensions by
Kao {\em et al.}\cite{Kao93}, who studied the Brownian motion of single 
particles at short times where the mean-square diplacement is not simply
proportional to $t$ \cite{Weitz92}.  As a result, $\mu(\omega,t)$ does not 
follow the linear time law of Eq.\ (\ref{3.43}).
Experimentalists prefer a different picture for DWS which they have developed 
for colloidal suspensions \cite{Weitz92}. They sum up all possible light 
paths in the scattering 
medium to arrive at the time-correlation function after some averaging
procedure. We will show here that our approach is totally equivalent to this 
picture. However it has the advantage that it automatically tells us how to
perform this averaging procedure.

DWS experiments are usually performed with continuous light sources and the 
diffusion equation reduces to
\begin{equation}
\label{3.44}
[ \mu(\omega,t) - \bbox{\nabla} \cdot 
\bbox{D}(\omega) \bbox{\nabla}] \,  W(\bbox{R},\,.\,,t) = 
\varrho(\bbox{R},\,.\,) \enspace.
\end{equation}
We can rewrite this equation as the Laplace transform of a problem,
where the source is a light pulse,
\begin{equation}
\label{3.45}
\Bigl[ \frac{\partial}{\partial \tau}  - \bbox{\nabla} \cdot 
\bbox{D}(\omega) \bbox{\nabla} \Bigr]\, P(\bbox{R},\tau) = 
\delta(\tau) \, \varrho(\bbox{R},\,.\,)
\end{equation}
with
\begin{equation}
\label{3.46}
W(\bbox{R},\,.\,,t) = \int_{-0}^{\infty} P(\bbox{R},\tau) \,
\text{exp}[-\mu(\omega,t) \tau ] \, d\tau \enspace.
\end{equation}
The lower limit $-0$ means a small negative time $\tau$ in order to pick up the
delta function. We can interpret $P(\bbox{R},\tau)$ (after an appropriate
normalization) as the probability that light, emitted by the source at time 
$\tau = 0$, arrives at the detector at point $\bbox{R}$ after a time $\tau$.
Then the time-correlation function $W(\bbox{R},\,.\,,t)$ follows after a
summation over all light paths where each light path contributes a
factor $\text{exp}[-\mu(\omega,t) \tau ]$ to the decay of 
$W(\bbox{R},\,.\,,t)$.
For isotropic systems $\tau$ is directly connected to the path length 
$s=c\tau/n$ of light and $P(\bbox{R},\tau)$ also represents the
path-length distribution. The exact form of $P(\bbox{R},\tau)$ depends on the
choice of the light source and the boundary conditions \cite{Weitz92} which
we don't address here.


\section{Light Diffusion in Nematic Liquid Crystals} \label{sec diff}

The nematic liquid crystalline phase consists of rodlike organic molecules
which tend to align parallel to each other but which show no long-range
positional order of their centers of mass. The local average direction of the
molecules is described by a unit vector $\bbox{n}(\bbox{r},t)$ called director.
It appears in the local dielectric tensor
\begin{equation}
\label{4.01}
\bbox{\varepsilon}(\bbox{r},t) = \varepsilon_{\perp} \bbox{1} +
\Delta \varepsilon [\bbox{n}(\bbox{r},t) \otimes \bbox{n}(\bbox{r},t)] 
\enspace,
\end{equation}
where $\varepsilon_{\perp}$ and $\varepsilon_{\|}$ are the dielectric constants
for electric fields, respectively, perpendicular and parallel to the director
and where 
$\Delta \varepsilon = \varepsilon_{\|} - \varepsilon_{\perp}$ stands for
the dielectric anisotropy. The energetically favored state of a nematic phase
is a uniform director field $\bbox{n}(\bbox{r},t)=\bbox{n}_0$ throughout the 
sample. Its distortion costs energy which can be calculated from the
Frank-Oseen-Z\"ocher free energy \cite{Gennes93}:
\begin{eqnarray}
F[\bbox{n}] & = & \frac{1}{2} \, \int [ K_1 (\text{div}\bbox{n})^2 
   + K_2 (\bbox{n} \cdot \text{curl}\bbox{n})^2 \nonumber \\
\label{4.02}
 & & + \, K_3 (\bbox{n} \times \text{curl}\bbox{n})^2 
    - \Delta \chi (\bbox{n} \cdot \bbox{H})^2] \, d^3r \enspace.
\end{eqnarray}
There are three characteristic distortions, called splay ($K_1$), twist
($K_2$) and bend ($K_3$), where $K_1$, $K_2$ and $K_3$ are the Frank elastic
constants. We also include a magnetic-field term with 
$\Delta \chi= \chi_{\|} - \chi_{\perp}$ the anisotropy of the magnetic
susceptibility. If $\Delta \chi > 0$ an alignment of the director parallel to
the field $\bbox{H}$ is favored. Even in a uniformly aligned sample
there exist thermally induced fluctuations of the director:
\begin{equation}
\label{4.03}
\bbox{n}(\bbox{r},t) = \bbox{n}_0 + \delta \bbox{n}(\bbox{r},t) \enspace .
\end{equation}
They lead to fluctuations in the local dielectric tensor and hence scatter
light. This is the physical phenomenon for which we want to calculate the
diffusion approximation of light. In order to apply the formulas of the 
preceeding 
section, we have to look first at the light propagation in an uniaxial media.
Then we need to calculate the structure factor, associated with the director 
fluctuations, which governs the single-light-scattering event.

\subsection{Light propagation in uniaxial media} \label{subsec diff.prop}

Let us, for a moment, suppress the fluctuations of the director and look at a
homogeneous medium with a uniaxial dielectric tensor
\begin{equation}
\label{4.1}
\bbox{\varepsilon}_0 = \varepsilon_{\perp} \bbox{1} +
\Delta \varepsilon [\bbox{n}_0 \otimes \bbox{n}_0] 
\enspace.
\end{equation}
The equilibrium value $\bbox{n}_0$ of the director is also called the
{\em optical axis\/} because it establishes a special axis for light 
propagation. The two light modes follow from solving the eigenvalue equation\
 (\ref{2.3}). They are well described in the literature \cite{Landau60}. We 
mainly summarize the results here, introduce some notation, and perform an
interesting variable transformation for later use.

One light mode is immediately obvious. Its polarization vectors 
$\bbox{e}_2(\hat{\bbox{k}})$ and $\bbox{d}^2(\hat{\bbox{k}})$ are both
perpendicular to $\bbox{n}_0$ and $\hat{\bbox{k}}$ with a refractive index 
$n_2 = \sqrt{\varepsilon_{\perp}}$. Since it behaves as in an isotropic
system, it is named the {\em ordinary\/} light mode. 
We choose $\bbox{n}_0$ parallel
to the $z$ axis and write $\hat{\bbox{k}}$ in spherical coordinates:
\begin{equation}
\label{4.2}
\bbox{n}_0 = \left[ \begin{array}{c} 
                        0 \\ 0 \\ 1
                     \end{array} \right] \quad 
\text{and} \quad
\hat{\bbox{k}} = \left[ \begin{array}{c}
                            \text{sin} \vartheta_{\bbox{k}} \,
                            \text{cos} \varphi_{\bbox{k}} \\
                            \text{sin} \vartheta_{\bbox{k}} \,
                            \text{sin} \varphi_{\bbox{k}} \\
                            \text{cos} \vartheta_{\bbox{k}}
                         \end{array} \right] \enspace.
\end{equation}
Then the ordinary light ray is represented by
\begin{equation}
\label{4.3}
n_2= \sqrt{\varepsilon_{\perp}} \quad \text{and} \quad
\bbox{e}_2(\hat{\bbox{k}}) = \frac{1}{n_2} \,
                         \left[ \begin{array}{c}
                           -\text{sin} \varphi_{\bbox{k}} \\
                            \text{cos} \varphi_{\bbox{k}} \\
                               0
                         \end{array} \right] \enspace.
\end{equation}

In the {\em extraordinary\/} light mode, the refractive
index depends on $\hat{\bbox{k}}$, and the polarization vector 
$\bbox{e}_1(\hat{\bbox{k}})$ is generally not perpendicular to
$\hat{\bbox{k}}$. It follows from $\bbox{e}_1(\hat{\bbox{k}}) = 
\bbox{\varepsilon}^{-1}_{0} \bbox{d}^1(\hat{\bbox{k}})$ where the polarization
vector $\bbox{d}^1(\hat{\bbox{k}})$ is determined by 
$\bbox{d}^1(\hat{\bbox{k}}) \perp \bbox{d}^2(\hat{\bbox{k}})$ and
$\bbox{d}^1(\hat{\bbox{k}}) \perp \hat{\bbox{k}}$. The refractive index 
can be calculated from the eigenvalue equation\ (\ref{2.3}) or 
Fresnel's equation\ (\ref{2.7}):
\begin{equation}
\label{4.4}
n_1(\hat{\bbox{k}}) = n_2 \widetilde{n}_1(\hat{\bbox{k}}) \,\enspace 
\text{with}
\enspace \, \widetilde{n}_1(\hat{\bbox{k}}) = \left[
 \frac{1+\alpha}{1+\alpha \text{cos}^2\vartheta_{\bbox{k}}} \right]^{1/2} ,
\end{equation}
where we have introduced the relative dielectric anisotropy of the system,
\begin{equation}
\label{4.5}
\alpha = \Delta \varepsilon / \varepsilon_{\perp} \enspace.
\end{equation}
Equation\ (\ref{4.4}) does not represent the most symmetric form for 
$n_1(\hat{\bbox{k}})$, but we found it useful for our calculations.
Finally, we get
\begin{equation}
\label{4.6}
\bbox{e}_1(\hat{\bbox{k}}) = \frac{\widetilde{n}_1(\hat{\bbox{k}})}{n_2}\, 
\left( \text{cos} \vartheta_{\bbox k} \,
  \left[ \begin{array}{c} \text{cos} \varphi_{\bbox{k}} \\
                          \text{sin} \varphi_{\bbox{k}} \\ 
                          0
         \end{array} \right]
       -\frac{\text{sin} \vartheta_{\bbox k}}{1+\alpha} \,
  \left[ \begin{array}{c} 0 \\ 0 \\ 1 \end{array} \right] \,\right) \enspace.
\end{equation}

In the structure factor, $\hat{\bbox{k}}$ always appears together with the
refraction index. For the extraordinary mode we, therefore, replace the
angular variables $\text{cos}\vartheta_{\bbox{k}}$ and 
$\text{sin}\vartheta_{\bbox{k}}$ by an equivalent set,
\begin{equation}
\label{4.7}
C_{\bbox{k}} := \widetilde{n}_1(\hat{\bbox{k}}) \, \text{cos}
\vartheta_{\bbox{k}} \quad \text{and} \quad
S_{\bbox{k}} := \widetilde{n}_1(\hat{\bbox{k}}) \, \text{sin}
\vartheta_{\bbox{k}} \enspace ,
\end{equation}
in which the refraction index takes the form
\begin{equation}
\label{4.9}
\widetilde{n}_1(\hat{\bbox{k}}) = \sqrt{1+\alpha(1-C^2_{\bbox{k}})} \enspace.
\end{equation}
The ``trigonometric'' identity,
\begin{equation}
\label{4.10}
S_{\bbox{k}}^2 + (1+\alpha) C_{\bbox{k}}^2 = 1+\alpha
\end{equation}
is valid. The new coordinate $C_{\bbox{k}}$ ranges from $-1$ to 1
and contains the same information as $\text{cos}\vartheta_{\bbox{k}}$. 
Thus, we can choose, e.g., spherical harmonics as basis functions on the unit 
sphere with
$\text{cos}\vartheta_{\bbox{k}}$ replaced by $C_{\bbox{k}}$. The usefulness of
$C_{\bbox{k}}$ appears when we calculate its differential with respect to 
$\text{cos} \vartheta_{\bbox{k}}$:
\begin{equation}
\label{4.11}
(1+\alpha)\,dC_{\bbox{k}} =
\widetilde{n}_1^3(\hat{\bbox{k}}) d\text{cos}\vartheta_{\bbox{k}} \enspace.
\end{equation}
The differential $dC_{\bbox{k}}$ absorbs the factor 
$\widetilde{n}_1^3(\hat{\bbox{k}})$ 
which always comes with our integrals thus making them easier:
\begin{eqnarray}
\int_{\hat{\bbox{k}}^{1}} \ldots \enspace & = &
\int \frac{d\Omega_{\bbox{k}}}{(2\pi)^3}
n_{1}^3(\hat{\bbox{k}}) \,\, \ldots \nonumber \\ 
\label{4.12}
 & = & n_2^3 (1+\alpha) 
\int \frac{dC_{\bbox{k}} d\varphi_{\bbox{k}}}{(2\pi)^3} \, \ldots \enspace.
\end{eqnarray}
Alternatively, we can say that we have constructed a complete set
of basis functions in the variables 
$\text{cos}\vartheta_{\bbox{k}}$ and $\varphi_{\bbox{k}}$
with respect to the weight function 
$\widetilde{n}_1(\hat{\bbox{k}})$.
In the following we will also use the notation $C_{\bbox{k}} = 
\text{cos} \vartheta_{\bbox{k}}$ and $S_{\bbox{k}} = 
\text{sin} \vartheta_{\bbox{k}}$ for ordinary light modes.

\subsection{Structure factor for director fluctuations} 
\label{subsec diff.struc}

\begin{figure}
\centerline{\psfig{figure=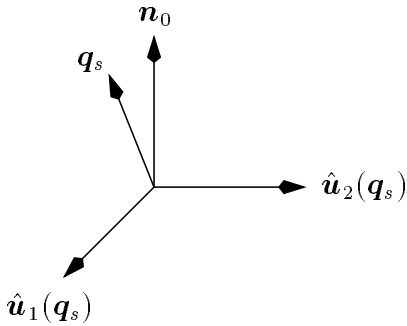}}

\vspace{.3cm}

\caption[]{Basis vectors $\hat{\bbox{u}}_1(\bbox{q}_s)$ and 
$\hat{\bbox{u}}_2(\bbox{q}_s)$ for given director $\bbox{n}_0$ and 
wave vector $\bbox{q}_s$.}
\label{fig3a}
\end{figure}

Here we will derive the structure factor for director fluctuations. To begin
we recall the general form of the time-dependent structure factor for light
scattering in an anisotropic dielectric,
\begin{eqnarray}
[\bbox{B}^{\omega}_{\bbox{k}^{\alpha}\bbox{q}^{\beta}}(t)]_{\alpha\beta}
 & = & \frac{\omega^4}{c^4} \, \langle \, 
\bbox{e}_{\alpha}(\hat{\bbox{k}}) \cdot \delta \bbox{\varepsilon}
(\bbox{q}_s,t) \bbox{e}_{\beta}(\hat{\bbox{q}}) \nonumber \\
\label{4.13}
 & & \qquad
[\bbox{e}_{\alpha}(\hat{\bbox{k}}) \cdot \delta \bbox{\varepsilon}
(\bbox{q}_s,0)\bbox{e}_{\beta}(\hat{\bbox{q}})]^{\ast} \,\rangle \enspace,
\end{eqnarray}
where $ \bbox{q}_s = \frac{\omega}{c} \, (n_{\alpha}\hat{\bbox{k}} 
 - n_{\beta}\hat{\bbox{q}}) $ stands for the scattering vector.
We first need the Fourier component $\delta \bbox{\varepsilon}
(\bbox{q}_s,t)$ of the fluctuating part of the dielectric tensor. We insert
$\bbox{n}(\bbox{r},t) = \bbox{n}_0 + \delta \bbox{n}(\bbox{r},t)$ into
$\bbox{\varepsilon}(\bbox{r},t)$ of Eq.\ (\ref{4.01}) and
collect the first-order terms in $\delta \bbox{n}(\bbox{r},t)$. After
a Fourier transformation we arrive at
\begin{equation}
\label{4.13a}
\delta \bbox{\varepsilon}(\bbox{q}_s,t) = \Delta\varepsilon \, [\bbox{n}_0 
 \otimes \delta \bbox{n}(\bbox{q}_s,t) + \delta \bbox{n}(\bbox{q}_s,t)
 \otimes \bbox{n}_0]
\end{equation}
with $\delta \bbox{n}(\bbox{q}_s,t)$ the amplitude of a director
mode. It has only two components since the director is a unit vector.
Furthermore, for small fluctuations, $\delta \bbox{n}$ is perpendicular to
$\bbox{n}_0$. An appropriate basis for $\delta \bbox{n}(\bbox{q}_s,t)$,
as shown in Fig. \ref{fig3a},
consists of a unit vector $\hat{\bbox{u}}_1(\bbox{q}_s)$, 
lying in the plane defined by $\bbox{n}_0$ and $\bbox{q}_s$, and a second 
one, $\hat{\bbox{u}}_2(\bbox{q}_s)$, perpendicular to this plane:
\begin{equation}
\label{4.14}
\delta \bbox{n}(\bbox{q}_s,t) = 
 \delta n_1(\bbox{q}_s,t) \hat{\bbox{u}}_1(\bbox{q}_s) + 
 \delta n_2(\bbox{q}_s,t) \hat{\bbox{u}}_2(\bbox{q}_s) \enspace.
\end{equation}
Next we need the temporal autocorrelation function of the 
director modes \cite{Gennes93}, which is rather complex. We state it and then
explain the individual terms:
\begin{eqnarray}
\label{4.15}
\lefteqn{ \langle \,\delta\bbox{n}(\bbox{q}_s,t) \otimes
\delta\bbox{n}^{\ast}(\bbox{q}_s,0) \,\rangle = }\quad \nonumber\\ & & 
\sum_{\delta =1}^{2} \frac{k_{\text{B}}T}{K_{\delta}(\bbox{q}_s)} \, 
\exp\left[-\frac{K_{\delta}(\bbox{q}_s)}{\eta_{\delta}(\bbox{q}_s)} \,t \right]
\, \hat{\bbox{u}}_{\delta}(\bbox{q}_s) 
 \otimes \hat{\bbox{u}}_{\delta}(\bbox{q}_s)
\end{eqnarray}
with
\begin{equation}
\label{4.16}
K_{\delta}(\bbox{q}_s)= K_{\delta} q_{\perp}^2 + K_3 q_{\|}^2 + \Delta
\chi H^2 \enspace,
\end{equation}
where $q_{\perp}$ and $q_{\|}$ are the components of $\bbox{q}_s$, 
perpendicular and parallel to $\bbox{n}_0$.
The chosen basis for $\delta \bbox{n}(\bbox{q}_s,t)$ provides the
normal coordinates,
because the correlation function is already diagonal. The free energy of
one director mode follows from the Frank-Oseen-Zocher free energy (\ref{4.02})
as $K_{\delta}(\bbox{q}_s) |\delta n_{\delta}(\bbox{q}_s,t)|^2$. For a general
vector $\bbox{q}_s$ it is either composed of splay and bend 
($\delta =1$) or twist and bend distortions ($\delta = 2$). The factor 
$k_{\text{B}}T / K_{\delta}(\bbox{q}_s)$ results from the equipartition
theorem giving the mean-square amplitude of each mode.  The dynamics of
the director modes is described by the Leslie-Erickson equations 
\cite{Chandrasekhar92}. They combine the Navier-Stokes equation for a 
uniaxial media with dynamical equations for the director. A detailed analysis 
\cite{Chandrasekhar92} shows that director modes are diffusive with a 
relaxation
frequency given by the quotient of elastic [$K_{\delta}(\bbox{q}_s)$] and
viscous [$\eta_{\delta}(\bbox{q}_s)$] forces. This is the origin of the
exponential factor in the correlation function. The viscosity 
$\eta_{\delta}(\bbox{q}_s)$ is a combination of several Leslie viscosities 
which we will address in subsection \ref{subsec diff.dws}.
We neglect a fast relaxing
part which comes from the coupling between the director and the fluid 
motion. Finally we are able to write down the structure factor
\begin{eqnarray}
\lefteqn{
[\bbox{B}^{\omega}_{\bbox{k}^{\alpha}\bbox{q}^{\beta}}(t)]_{\alpha \beta} 
=  (\Delta \varepsilon)^2 
k_{\text{B}}T \, \frac{\omega^4}{c^4\rule[-5mm]{0mm}{3mm}} \qquad }
 \nonumber \\
\label{4.17}
 & & \qquad \times \sum_{\delta=1}^{2}
\frac{N(\bbox{e}_{\alpha},\bbox{e}_{\beta},\hat{\bbox{u}}_{\delta})}{
 K_{\delta}(\bbox{q}_s)} \,
\exp\biggl[   - \frac{K_{\delta}(\bbox{q}_s)}{\eta_{\delta}(\bbox{q}_s)}
 \, \, t \,\biggr]
\end{eqnarray}
with
\begin{equation}
\label{4.17a}
N(\bbox{e}_{\alpha},\bbox{e}_{\beta},\hat{\bbox{u}}_{\delta}) =
[(\bbox{n}_0\cdot \bbox{e}_{\beta})
(\hat{\bbox{u}}_{\delta} \cdot \bbox{e}_{\alpha})
 + (\hat{\bbox{u}}_{\delta} \cdot \bbox{e}_{\beta})
 (\bbox{n}_0 \cdot \bbox{e}_{\alpha})
]^2
\end{equation}
a geometrical factor. It has two interesting implications. First, there
exists no scattering of an ordinary light ray into an ordinary light ray because 
$\bbox{e}_{2} \perp \bbox{n}_0$, and therefore $N=0$. Second, forward and 
backward scattering along $\bbox{n}_0$ is always forbidden. The other terms
of $[\bbox{B}^{\omega}_{\bbox{k}^{\alpha}\bbox{q}^{\beta}}(t)]_{\alpha \beta}$
are familiar from the previous discussion. The scattering mean-free-path
$l_{\alpha}(\hat{\bbox{k}},\omega)$, defined in Eq.\ (\ref{3.14}), has been
already discussed in detail by two groups \cite{Val'kov86,Langevin75}. Its
dependence on $\hat{\bbox{k}}$ and polarization was calculated by
Val'kov and Romanov \cite{Val'kov86}. We just stress one point.
The structure factor diverges for $H=0$ and $\bbox{q}_s \rightarrow \bbox{0}$,
and we expect the scattering mean-free-paths to be zero in an infinite
medium. But $\bbox{q}_s =\bbox{0}$ can only occur in scattering events
where the extraordinary polarization is preserved. Hence, only the scattering
mean-free-path $l_1(\hat{\bbox{k}},\omega)$ of an extraordinary light-ray is 
zero. In Appendix \ref{sec struc} we give 
$[\bbox{B}^{\omega}_{\bbox{k}^{1}\bbox{q}^{1}}(0)]_{11}$ for 
$\bbox{q}^{1} \rightarrow \bbox{k}^{1}$ (the notation we use there is explained below).
From this form it is obvious that $l_1^{-1}(\hat{\bbox{k}},\omega)$
diverges weakly like 
$-\text{ln}[\Delta \chi H^2/(K_3 n_2^2\omega^2/c^2)]$
for magnetic fields much smaller than $\sqrt{K_3/\Delta \chi} n_2\omega/c$.
The explicit form of the structure factor is very complex because it
explicitly depends on the direction of the incoming ($\hat{\bbox{k}}$) and 
outgoing ($\hat{\bbox{q}}$) light through the geometrical factor. We mention
some symmetry properties which facilitate its handling. The first two
are quite obvious:
\begin{equation}
\label{17b}
[\bbox{B}^{\omega}_{\bbox{k}^{\alpha}\bbox{q}^{\beta}}(0)]_{\alpha \beta}
= [\bbox{B}^{\omega}_{\bbox{q}^{\beta} \bbox{k}^{\alpha}}(0)]_{\beta \alpha}
\end{equation}
and
\begin{equation}
\label{17c}
[\bbox{B}^{\omega}_{\bbox{k}^{\alpha}\bbox{q}^{\beta}}(0)]_{\alpha \beta}
= [\bbox{B}^{\omega}_{-\bbox{k}^{\alpha}-\bbox{q}^{\beta}}(0)]_{\alpha \beta}
\enspace.
\end{equation}
The structure factor has to reflect the symmetry of the nematic phase described
by the group $D_{\infty h}$. For example, it has to exhibit the rotational 
symmetry about the director $\bbox{n}_0$. If we choose
\begin{equation}
\label{4.18}
\hat{\bbox{k}} = \left[ \begin{array}{c}
                            \text{sin} \vartheta_{\bbox{k}} \,
                            \text{cos} \varphi_{\bbox{k}} \\
                            \text{sin} \vartheta_{\bbox{k}} \,
                            \text{sin} \varphi_{\bbox{k}} \\
                            \text{cos} \vartheta_{\bbox{k}}
                         \end{array} \right]
\,\, \text{and} \enspace
\hat{\bbox{q}} = \left[ \begin{array}{c}
                            \text{sin} \vartheta_{\bbox{q}} \,
                            \text{cos} \varphi_{\bbox{q}} \\
                            \text{sin} \vartheta_{\bbox{q}} \,
                            \text{sin} \varphi_{\bbox{q}} \\
                            \text{cos} \vartheta_{\bbox{q}}
                         \end{array} \right] \, ,
\end{equation}
then 
$[\bbox{B}^{\omega}_{\bbox{k}^{\alpha}\bbox{q}^{\beta}}(0)]_{\alpha \beta}$ 
depends only on the relative azimuthal angle 
$\varphi = \varphi_{\bbox{q}} - \varphi_{\bbox{k}}$. The existence of mirror 
planes, containing $\bbox{n}_0$, implies that the structure factor should 
be invariant under $\varphi \rightarrow -\varphi$. In appendix 
\ref{sec struc} we give its explicit form in terms of 
$C_{\bbox{k}}$, $S_{\bbox{k}}$, $C_{\bbox{q}}$, $S_{\bbox{q}}$ and $\varphi$.
We also introduce a scaled structure factor
$[\widetilde{\bbox{B}}^{\omega}_{\bbox{k}^{\alpha}\bbox{q}^{\beta}}(0)]_{
 \alpha \beta}$ through
\begin{equation}
\label{4.19}
[\bbox{B}^{\omega}_{\bbox{k}^{\alpha}\bbox{q}^{\beta}}(0)]_{\alpha \beta}
= (\Delta \varepsilon)^2 \, \frac{\omega^2}{c^2} \,
  \frac{k_{\text{B}}T}{K_3 n_2^6} \, 
[\widetilde{\bbox{B}}^{\omega}_{\bbox{k}^{\alpha}\bbox{q}^{\beta}}(0)]_{
 \alpha \beta} \enspace ,
\end{equation}
which depends on scaled parameters: the relative dielectric anisotropy 
$\alpha = \Delta \varepsilon/\varepsilon_{\perp}$, the Frank elastic constants
\begin{equation}
\label{4.20}
\overline{K}_1 = K_1/K_3 \quad \text{and} \quad \overline{K}_2 = K_2/K_3
\enspace,
\end{equation}
and the magnetic field
\begin{equation}
\label{4.21}
h=H/H_0 \quad \text{with} \quad H_0 = n_2\,\frac{\omega}{c} \,
\sqrt{\frac{K_3}{\Delta \chi}}  \enspace.
\end{equation}
If we introduce the magnetic coherence length
\begin{equation}
\label{4.22}
\xi_3 = \sqrt{ \frac{K_3}{\Delta \chi H^2} } \enspace,
\end{equation}
which gives the length scale over which director fluctuations are correlated,
we obtain for the scaled magnetic field
\begin{equation}
\label{4.22b}
h = \frac{\lambda}{2\pi \xi_3}  \enspace,
\end{equation}
where $\lambda = n_2\omega/c$. Thus, $h=1$ corresponds to a very short
coherence length of $\lambda/2\pi$.

In the next subsection we will use the material parameters of a 
typical nematic compound CB5 to discuss the diffusion constants. 5CB is
liquid crystalline at room temperature, and we use the parameters for 5$\,$K
below the nematic-isotropic transition \cite{Collings95}. 
The bend elastic constant is
$K_3=5.3 \times 10^{-7} \,\text{dyne}$ and is, 
as usual for conventional thermotropic nematic liquid
crystals, larger then the splay and twist constants: 
$\overline{K}_1 = 0.79$ and $\overline{K}_2 = 0.43$. For green light 
($\omega/c = 1.15 \times 10^5 \, \text{cm}^{-1}$) the dielectric constant is
$\varepsilon_{\perp}= 2.381$ and the anisotropy is $\alpha=0.228$ corresponding
to the refractive indices $n_1(\vartheta=90^{\circ}) = 1.710$ and $n_2=1.543$.
Finally, the magnetic anisotropy is $\Delta \chi = 0.95 \times 10^{-7}$,
from which we obtain a characteristic magnetic field of 
$H_0 = 3.6 \times 10^5\, \text{G}$.

\subsection{Light diffusion -- results} \label{subsec diff.res}

In this subsection we discuss the two essential components 
$D_{\|}$ and $D_{\perp}$ of the diffusion tensor $\bbox{D}(\omega)$,
which describe the diffusion of light, respectively, parallel and perpendicular
to the director $\bbox{n}_0$. Before we can apply our general formulas from
Eqs.\ (\ref{3.31}), (\ref{3.32}) and (\ref{3.39}) we need to be more
specific about our basis functions $\bbox{\varphi}_i(\hat{\bbox{k}})$, and
we need the derivatives of the components $[\bbox{G}_0^{-1}]_{\alpha}$ which 
we gave in Eq.\ (\ref{2.11}). The calculations for the latter are 
straightforward and the results for the extraordinary and ordinary light mode 
are, respectively,
\begin{eqnarray}
\lefteqn{\left. \widetilde{n}_1(\hat{\bbox{k}}) \,
 \frac{\partial [\bbox{G}_0^{-1}]_{1}}{\partial \bbox{k}}
 \right|_{\hat{\bbox{k}}}\!\! \cdot \bbox{K} \qquad \qquad} \nonumber \\
\label{4.23}
 & & \qquad \qquad =\, - \frac{2}{\varepsilon_{\perp}}
  \, \Bigl( \frac{\sqrt{1- C_{\bbox{k}}^2}\text{cos} 
  \varphi_{\bbox{k}}}{\sqrt{1+\alpha}} K_{\perp}
  + C_{\bbox{k}} K_{\|}\Bigr)
\end{eqnarray}
and
\begin{equation}
\label{4.24}
\left. \frac{\partial [\bbox{G}_0^{-1}]_{2}}{\partial \bbox{k}}
\right|_{\hat{\bbox{k}}}\!\! \cdot \bbox{K}
= -\frac{2}{\varepsilon_{\perp}} \, (\sqrt{1-C_{\bbox{k}}^2} 
 \text{cos} \varphi_{\bbox{k}} K_{\perp} + C_{\bbox{k}} K_{\|} ) \enspace.
\end{equation}
In the above, we chose
\begin{equation}
\label{4.25}
\bbox{K} = [ K_{\perp}, 0 , K_{\|} ] \enspace,
\end{equation}
wrote $\hat{\bbox{k}}$ in spherical coordinates as in Eq.\ (\ref{4.18}) and
then switched to the new $C$ coordinate. The right-hand side of 
Eq.\ (\ref{4.24}) for the ordinary mode is the isotropic result 
$\hat{\bbox{k}}\cdot \bbox{K}$, which is modified in the extraordinary case.
As our basis functions, we choose
\begin{equation}
\label{4.26}
\varphi^{\alpha}_{i}(\hat{\bbox{k}}) \enspace \longrightarrow \enspace
\varphi^{\alpha}_{\gamma l m}(\hat{\bbox{k}}) = \delta^{\alpha}_{\gamma} \,
\varphi_{lm}(C_{\bbox{k}}, \varphi_{\bbox{k}}) \enspace ,
\end{equation}
where $\varphi_{lm}(C_{\bbox{k}}, \varphi_{\bbox{k}})$ stands for a real
combination of the spherical harmonics 
$Y_{lm}(C_{\bbox{k}}, \varphi_{\bbox{k}})$ and
$Y_{l-m}(C_{\bbox{k}}, \varphi_{\bbox{k}})$ in the new coordinate
$C_{\bbox{k}}$ instead of $\text{cos} \vartheta_{\bbox{k}}$.
It will soon become clear why the coordinate $C_{\bbox{k}}$ is so helpful.
The index $\gamma$ stands for the basis of the tensor space, which we
identify here with our basis $\bbox{e}_{\alpha} \otimes \bbox{e}_{\alpha}$. 
Only functions with odd parity, i.e., with odd $l$, contribute to the 
diffusion. This is 
immediately obvious from the parity of the structure factor and the derivative
of $[\bbox{G}_0^{-1}]_{\alpha}$. Furthermore, we only need functions
$\varphi_{lm}(C_{\bbox{k}}, \varphi_{\bbox{k}})$ containing 
$\text{cos} m \varphi_{\bbox{k}}$ ($m \ge 0$);  
$\text{sin} m \varphi_{\bbox{k}}$ is not necessary because of the symmetry of 
$[\bbox{B}^{\omega}_{\bbox{k}^{\alpha}\bbox{q}^{\beta}}(0)]_{\alpha \beta}$.
With this choice of basis functions the matrix $\cal B$ is decomposed
into submatrices for each $m$, $[\cal B]^{\gamma lm}_{\gamma' l'm'} \propto
\delta_{m m'}$, because a term $\text{cos} m \varphi_{\bbox{k}}\,
\text{cos} n \varphi_{\bbox{k}}$, ($n \ne m$), is not compatible with
the rotational symmetry of the structure factor. Then, the
diffusion constants $D_{\|}$ and $D_{\perp}$ are, respectively, related to
$m=0$ and $m=1$:
\begin{eqnarray}
\label{4.27}
D_{\|} K_{\|}^2 & = &\frac{c}{2\overline{n^3}}
                    \sum_{\gamma\gamma' l l'}
                    [{\cal G}(\bbox{K})]_{\gamma l 0} \,
                    [{\cal B}^{-1}]^{\gamma l 0}_{\gamma' l' 0} \,
                    [{\cal G}(\bbox{K})]_{\gamma' l' 0} \\
\label{4.28}
D_{\perp} K_{\perp}^2 & = & \frac{c}{2\overline{n^3}}
                 \sum_{\gamma\gamma' l l'}
                 [{\cal G}(\bbox{K})]_{\gamma l 1} \,
                 [{\cal B}^{-1}]^{\gamma l 1}_{\gamma' l' 1} \,
                 [{\cal G}(\bbox{K})]_{\gamma' l' 1}
\end{eqnarray}
with
\begin{equation}
\label{4.29}
\overline{n^3} = n_2^3 (1+\alpha/2) = (\varepsilon_{\perp})^{3/2}
(1+\alpha/2) \enspace.
\end{equation}
Only elements of ${\cal G}(\bbox{K})$ with $l=1$ are non-zero,
$[{\cal G}(\bbox{K})]_{\gamma 1 m} \ne 0$, since the right-hand sides of
Eqs.\ (\ref{4.23}) and (\ref{4.24}) depend only on $l=1$ basis functions.
This rises the question of how 
important higher $l$ contributions are. For our calculations we,
therefore, choose $l=1$ and $l=3$ functions:
\begin{mathletters}
\label{4.30}
\begin{eqnarray}
\varphi_{10}(C, \varphi) & = &
\sqrt{6}\pi \, C \\
\varphi_{11}(C, \varphi) & = & 
\sqrt{6}\pi \, \sqrt{1-C^2} \, \text{cos} \varphi \\
\varphi_{30}(C, \varphi) & = & \textstyle 
\sqrt{\frac{7}{2}}\pi \, C(5C^2 -3) \\
\varphi_{31}(C, \varphi) & = &\textstyle 
\frac{\sqrt{21}}{2}\pi \, \sqrt{1-C^2}\, (1-5C^2) \,
\text{cos} \varphi
\end{eqnarray}
\end{mathletters}
$\!\!$with the normalization
\begin{equation}
\label{4.31}
\int \frac{dC d \varphi}{(2\pi)^3}
\varphi_{lm}(C, \varphi) \, 
\varphi_{l'm'}(C, \varphi) = \delta_{ll'} \delta_{mm'}
\enspace .
\end{equation}
The non-zero components  of ${\cal G}(\bbox{K})$ turn out to be
\begin{mathletters}
\label{4.32}
\begin{eqnarray}
[{\cal G}(\bbox{K})]_{110} & = & -2 \varepsilon_{\perp} (1+\alpha) K_{\|}
 /\sqrt{6} \\ \protect
[{\cal G}(\bbox{K})]_{210} & = & -2 \varepsilon_{\perp}  K_{\|} / \sqrt{6} \\
\protect
[{\cal G}(\bbox{K})]_{111} & = & -2 \varepsilon_{\perp} \sqrt{1+\alpha} 
K_{\perp} / \sqrt{6} \\ \protect
[{\cal G}(\bbox{K})]_{211} & = & -2 \varepsilon_{\perp}  K_{\perp} / \sqrt{6}
\enspace .
\end{eqnarray}
\end{mathletters}
$\!\!$The integrations to obtain the matrix elements of $\cal B$ requires more 
effort. We were able to calculate the integrals over $\varphi$ analytically 
with the help of
\begin{mathletters}
\label{4.33}
\begin{eqnarray}
\int d\varphi_{\bbox{k}} d\varphi_{\bbox{q}} 
 f(\varphi_{\bbox{q}}- \varphi_{\bbox{k}}) & = & 2\pi \int d \varphi
 f(\varphi) \\
\int d\varphi_{\bbox{k}} d\varphi_{\bbox{q}} 
 f(\varphi) 
 \text{cos}\varphi_{\bbox{k}} \, \text{cos}\varphi_{\bbox{q}} & = &
 \pi \int d \varphi f(\varphi) \text{cos}\varphi \\
\int d\varphi_{\bbox{k}} d\varphi_{\bbox{q}} 
 f(\varphi) 
 \text{cos}^2 \varphi_{\bbox{k}} & = &
 \pi \int d \varphi f(\varphi) \enspace .
\end{eqnarray}
\end{mathletters}
$\!\!$The results are listed in appendix \ref{sec struc}. The remaining 
integrations were performed numerically.

To discuss the diffusion constants we collect the prefactors of
the quantities involved to obtain an ``averaged'' transport mean-free-path 
\begin{equation}
\label{4.34}
l_0^{\ast} = 9\pi \, \frac{c_{\perp}^2}{\omega^2} \, 
\frac{K_3}{k_{\text{B}}T} \, \frac{1}{\alpha^2}
\end{equation}
and write the diffusion constant $D_{\|}$ and $D_{\perp}$ in the form
\begin{equation}
\label{4.35}
D_{\|} =  c_{\perp} l_0^{\ast} \widetilde{D}_{\|} / 3 
\quad ,\quad D_{\perp} = c_{\perp} l_0^{\ast} \widetilde{D}_{\perp} / 3
\enspace,
\end{equation}
reminiscent of isotropic systems. The numerical factors $\widetilde{D}_{\|}$
and $\widetilde{D}_{\perp}$ only depend on the scaled parameters $\alpha$, $h$,
$\overline{K}_1$, and $\overline{K}_2$, and $c_{\perp}$ is the speed of light
of the ordinary light ray. We stress that $l_0^{\ast}$ is an averaged
quantity and that our theory does not give a procedure to construct
transport mean-free-paths for different light directions (see however 
Ref. \cite{Jester96}).
The factor $9\pi$ in $l_0^{\ast}$ is chosen such that $\widetilde{D}_{\|}$
and $\widetilde{D}_{\perp}$ are approximately 1 in the limit of
an ``isotropic'' nematic with $\overline{K}_1=\overline{K}_2 =1$, 
$\alpha=0$, and $h=0$. We find $\widetilde{D}_{\|}=1.053$ and 
$\widetilde{D}_{\perp}=0.998$ with a small anisotropy of 
$D_{\|}/D_{\perp} = 1.06$ because of the inherent 
anisotropy in the nematic structure factor, which is represented by the 
geometrical factor 
$N(\bbox{e}_{\alpha},\bbox{e}_{\beta},\hat{\bbox{u}}_{\delta})$ of Eq.\ 
(\ref{4.17a}). We, at least qualitatively, understand why $D_{\|}$ is 
larger than $D_{\perp}$. The diffusion constants grow when the system's
ability to scatter light decreases. From equations (\ref{4.27}) and
(\ref{4.28}) for $D_{\|}$ and $D_{\perp}$, we recognize that 
the diffusion constants are, respectively, determined by scattering around 
the director ($m=0$) or perpendicular to it ($m=1$). However, forward
and backward scattering along the director is suppressed by the geometrical 
factor and we expect $D_{\|}$ to be larger than $D_{\perp}$. In a completely
isotropic system, where we set 
$N(\bbox{e}_{\alpha},\bbox{e}_{\beta},\hat{\bbox{u}}_{\delta})$ 
equal to 1, the transport mean-free-path is easy to
calculate. It is a factor of $4/9$ smaller than $l_0^{\ast}$, which again
demonstrates the effect of the geometrical factor.
For temperature $T=300\,\text{K}$ and the parameters of the compound
5CB, which we summarized in the last subsection, we obtain
$l_0^{\ast}=2.3\,\text{mm}$ in agreement with experiments 
\cite{Kao96,Jester96}.

\begin{figure}
\centerline{\psfig{figure=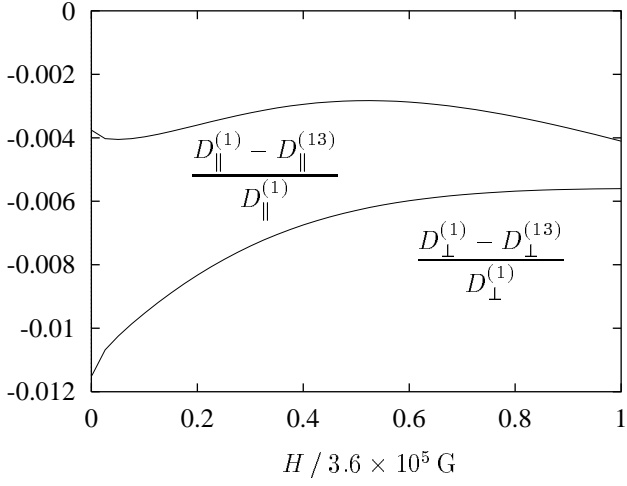}}

\vspace*{.3cm}

\caption[]{Field dependence of the relative changes of diffusion constants 
after $l=3$ spherical harmonics (superscript 13) are included in addition
to $l=1$ spherical harmonics (superscript 1).
Parameters are $\overline{K}_1=\overline{K}_2=1$ and $\alpha=0$. The magnetic
field is given relative to the characteristic field 
$H_0=3.6 \times 10^5\,\text{G}$ of the nematic compound 5CB.}
\label{fig3}
\end{figure}

\begin{figure}
\centerline{\psfig{figure=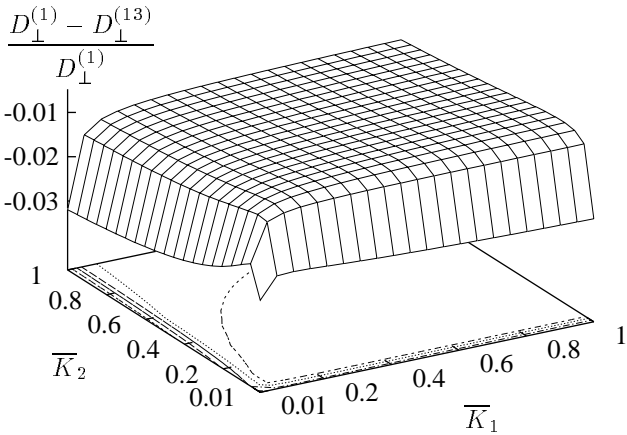}}

\vspace*{.3cm}

\caption[]{Relative difference as a function of $\overline{K}_1$ and
$\overline{K}_2$ at $\alpha=0$, $h=0.1$ between $D_{\perp}^{(1)}$ calculated
with $l=1$ spherical harmonics only and $D_{\perp}^{(13)}$ calculated with
both $l=1$ and $l=3$.}
\label{fig4}
\end{figure}

In the following we will explore the dependence of the diffusion constants
on the scaled parameters $\overline{K}_1$, $\overline{K}_2$, $\alpha$, and
$h$. In Figs.\ \ref{fig3} and \ref{fig4} we plot the relative changes of the 
diffusion constants when spherical harmonics of $l=3$, in addition to $l=1$, 
are included in the calculations. The field dependence in Fig.\ \ref{fig3}
shows that the changes are around 1\,\% or smaller and that $D_{\perp}$ is
more strongly affected by higher spherical harmonics. The same is valid for
the $\overline{K}_1$ and $\overline{K}_2$ dependence in Fig.\ \ref{fig4}.
Only for extreme situations $\overline{K}_1 < 0.1$ or $\overline{K}_2 < 0.1$
do the changes grow to 3\,\%. We conclude that the restriction to spherical
harmonics of $l=1$ gives a good approximation for the diffusion constants.
The following graphs will, however, all be presented with the $l=3$ 
contributions included.

\begin{figure}
\centerline{\psfig{figure=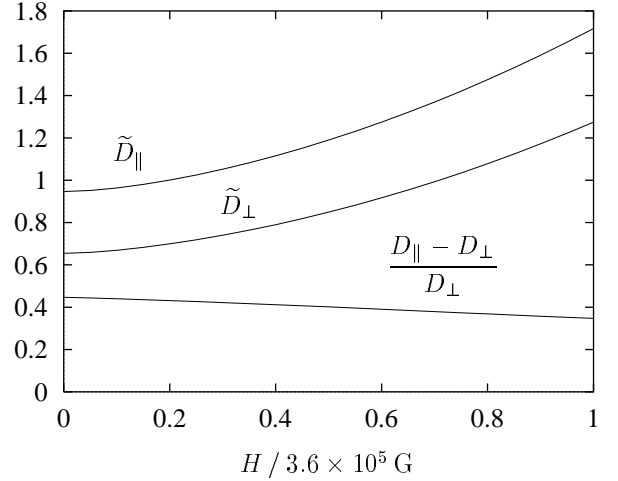}}

\vspace{.3cm}

\caption[]{Field dependence of the diffusion constants 
$\widetilde{D}_{\|}$ and $\widetilde{D}_{\perp}$ and the relative anisotropy
$(D_{\|} - D_{\perp})/D_{\perp}$ for the nematic compound 5CB 
($\overline{K}_1 = 0.79$, $\overline{K}_2 = 0.43$, $\alpha=0.228$ and
$H_0 = 3.6 \times 10^5 \text{G}$).}
\label{fig5}
\end{figure}

For the nematic compound 5CB we show in Fig.\ \ref{fig5} how the diffusion 
constants $\widetilde{D}_{\|}$ and $\widetilde{D}_{\perp}$ and the relative 
anisotropy $(D_{\|} - D_{\perp})/D_{\perp}$ behave in a magnetic field.
$\widetilde{D}_{\|}$ and $\widetilde{D}_{\perp}$ grow with $H$
because the magnetic field suppresses director fluctuations. The field
dependence of the relative anisotropy in the diffusion is weak. 
For ordinary magnetic fields up to $5\times10^4\,\text{G}$, which 
corresponds to a magnetic coherence
length $\xi_3$ of approximately $1 \,\mu\text{m}$, the changes in 
$\widetilde{D}_{\|}$ and $\widetilde{D}_{\perp}$ are small. The values for
$H =0$ read $\widetilde{D}_{\|} = 0.95$ and $\widetilde{D}_{\perp}= 0.65$
with a ratio $\widetilde{D}_{\|} / \widetilde{D}_{\perp} = 1.45 $ in good 
agreement with experiments \cite{Kao96,Jester96}. The reason
why we plot $\widetilde{D}_{\|}$, $\widetilde{D}_{\perp}$ and 
$(D_{\|} - D_{\perp})/D_{\perp}$ on a large field range is to show that
the quantities smoothly approach finite values at $h=0$. This is not
obvious since the structure factor possesses a singularity at $h=0$ and for
vanishing scattering vector $\bbox{q}_s$. Strictly speaking, our 
weak-scattering approximation is not applicable here. However, in a completely
isotropic model we understand why the quantities are finite. The familiar
factor $1- \text{cos} \vartheta_s$ in formula\ (\ref{3.41}) cancels the 
singularity from $\bbox{q}_s^2 \propto 1- \text{cos} \vartheta_s$.

\begin{figure}
\centerline{\psfig{figure=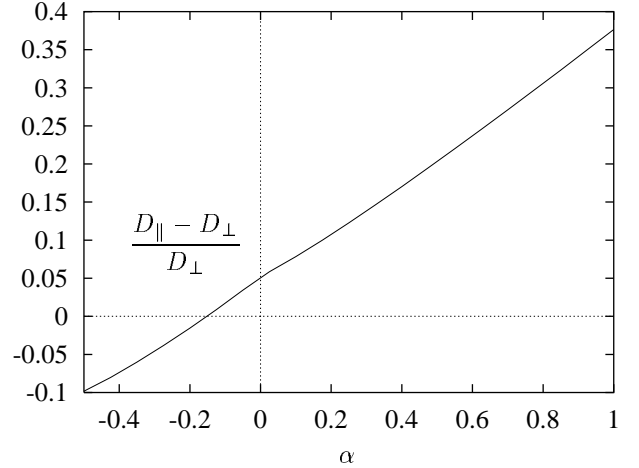}}

\vspace{.3cm}

\caption[]{Relative anisotropy $(D_{\|} - D_{\perp})/D_{\perp}$ depending
on the dielectric anisotropy $\alpha$ for $\overline{K}_1 = \overline{K}_2=1$
and $h=0.01$.}
\label{fig6}
\end{figure}

In Fig.\ \ref{fig6} we explore the anisotropy in the diffusion depending
on the anisotropy $\alpha$ in the dielectric constants. As already discussed,
when $\alpha=0$ there is a small non-zero value of 
$(D_{\|} - D_{\perp})/D_{\perp}$. This
grows with $\alpha$ because the speed of light of the extraordinary 
light ray is larger along the director than perpendicular to it. On the other
hand, for $\alpha < -0.15$, the anisotropy $(D_{\|} - D_{\perp})/D_{\perp}$
changes sign, and light diffuses faster perpendicular to the director.
This effect and the inversion point $D_{\|} = D_{\perp}$ should
be observable in discotic nematics where $\alpha$ is negative.


\vspace*{.2cm}

\vspace*{.1cm}

\begin{figure}
\centerline{\psfig{figure=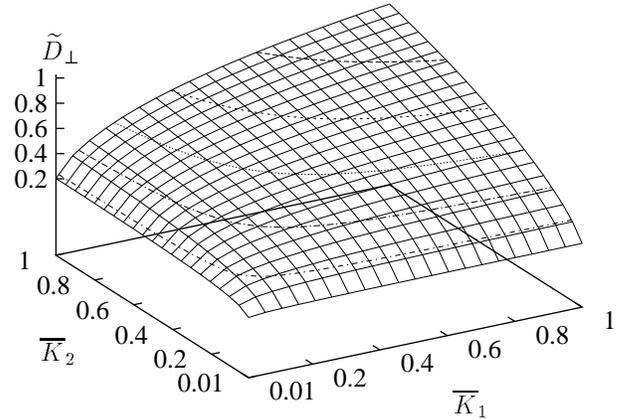}}

\vspace*{.3cm}

\caption[]{Diffusion constant $\widetilde{D}_{\perp}$ depending on
$\overline{K}_1$ and $\overline{K}_2$. Parameters are $\alpha=0$ and $h=0.1$.}
\label{fig7}
\end{figure}

\begin{figure}
\centerline{\psfig{figure=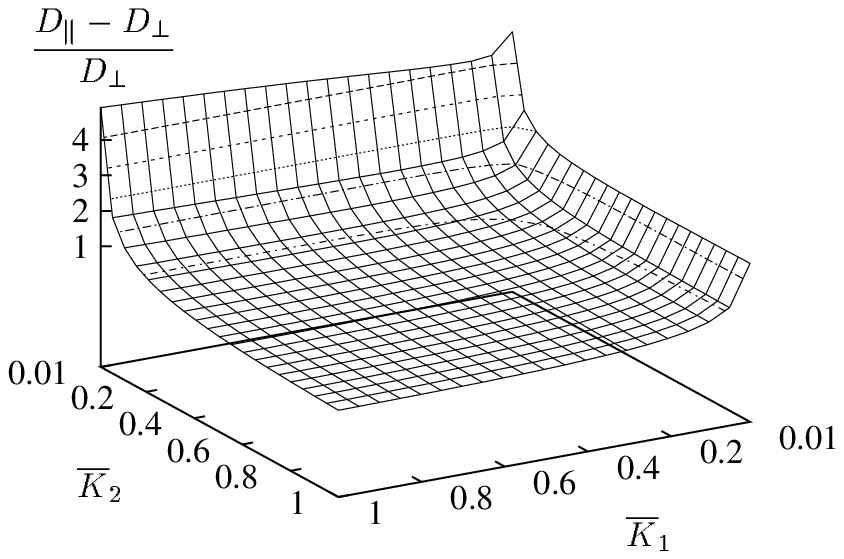}}

\vspace*{.3cm}

\caption[]{Relative anisotropy $(D_{\|} - D_{\perp})/D_{\perp}$ depending
on $\overline{K}_1$ and $\overline{K}_2$, which range from 0.01 to 1.
Parameters are $\alpha=0$ and $h=0.1$.}
\label{fig8}
\end{figure}

\begin{figure}
\centerline{\psfig{figure=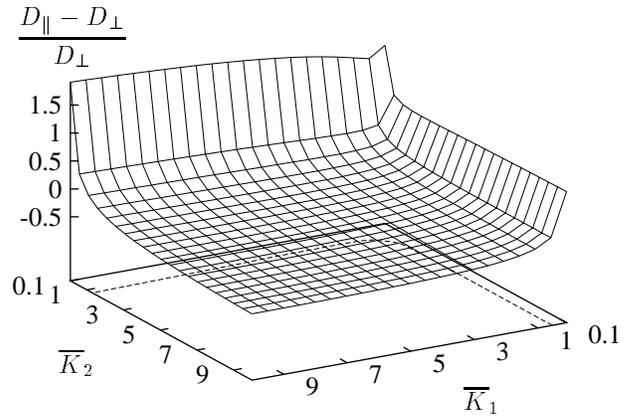}}

\vspace*{.3cm}

\caption[]{Relative anisotropy $(D_{\|} - D_{\perp})/D_{\perp}$ depending
on $\overline{K}_1$ and $\overline{K}_2$, which range from 0.1 to 10.
Parameters are $\alpha=0$ and $h=0.1$.}
\label{fig9}
\end{figure}

Finally, we discuss the dependence of the diffusion on the elastic constants
$\overline{K}_1$ and $\overline{K}_2$. In Fig.\ \ref{fig7} we show that
$\widetilde{D}_{\perp}$ decreases with the elastic constants since the
light scattering from the director modes increases. At the extreme values
$\overline{K}_1=\overline{K}_2=0.01$, we find $\widetilde{D}_{\perp}=0.07$.
The contour lines reveal an asymmetry between the splay 
($\overline{K}_1$) and the twist ($\overline{K}_2$) distortions. 
$\widetilde{D}_{\perp}$ decreases more strongly with $\overline{K}_2$.
The diffusion constant $\widetilde{D}_{\|}$ shows a similar behavior.
Fig.\ \ref{fig8} gives the  anisotropy $(D_{\|} - D_{\perp})/D_{\perp}$ for the
same range. It grows with decreasing elastic constants showing that
$\widetilde{D}_{\perp}$ is more affected by splay and twist distortions than
$\widetilde{D}_{\|}$.
The asymmetry between splay and twist is clearly visible. The last two graphs
cover the range of conventional thermotropic nematics where usually 
$\overline{K}_1 < 1$ and $\overline{K}_2 < 1$. In Fig.\ \ref{fig9} we extend 
this range to $\overline{K}_1 = \overline{K}_2 = 10$ and observe that the
anisotropy $(D_{\|} - D_{\perp})/D_{\perp}$ changes sign. The contour line
on the base of the coordinate system indicates where $D_{\|} = D_{\perp}$.
Roughly speaking, $D_{\|} < D_{\perp}$ if $\overline{K}_1 > 0.6$ and 
$\overline{K}_2 > 1.4$.
In a smectic-$A$ phase (SmA) twist and bend deformations are expelled
by the layered structure \cite{Gennes69b}, hence $\overline{K}_1 \ll 1$.
Unfortunately, this also means that 
certain scattering vectors show very weak scattering so that the diffusion
approximation cannot be achieved for reasonably sized samples. However, 
in the vicinity of a SmA-nematic phase transition, where the layered structure
softens (SmA) or starts to form (nematic phase), the diffusion approximation 
of light could be used to study the behavior of the Frank elastic constants 
close to the transition.
A second interesting system is a polymer nematic liquid crystal.
For long rigid rods one expects a large splay constant \cite{Meyer82,Donald92}.
Taratuta {\em et al.\/} \cite{Taratuta85} determined the Frank elastic 
constants for a special system and found the ratios $\overline{K}_1 = 0.85$
and $\overline{K}_2 = 0.07$ with an absolute value of $K_3 = 4.7 \times
10^{-7}\,\text{dynes}$, which is suitable for the diffusion approximation 
of light. From these parameters we predict a ``large'' ratio of 
$D_{\|} / D_{\perp} = 2.8$. The reported system has a very low dielectric
anisotropy $\alpha$ and scatters light only weakly. However, it should be 
possible to find systems which are more favorable regarding $\alpha$.

We also calculated a Taylor expansion for the relative anisotropy 
$(D_{\|} - D_{\perp})/D_{\perp}$ around $\overline{K}_1=\overline{K}_2=1$ and
$\alpha =0$ and found
\begin{eqnarray}
\frac{D_{\|} - D_{\perp}}{D_{\perp}} & \approx & 0.06 - 0.1
(\overline{K}_1 - 1) - 0.3 (\overline{K}_2 - 1) + 0.3 \alpha \nonumber \\
\label{4.36}
 & & + \, 0.1 (\overline{K}_1 - 1)^2 + 0.2 (\overline{K}_2 - 1)^2 \\
 & & - 0.08 (\overline{K}_1 - 1) (\overline{K}_2 - 1) \enspace,  \nonumber
\end{eqnarray}
where the coefficients are material independent.
Second order terms in $\alpha$ and couplings to $\overline{K}_1$ and 
$\overline{K}_2$ are negligible. The expansion summarizes the whole
discussion.

\subsection{Diffusing Wave Spectroscopy} \label{subsec diff.dws}

With Diffusing Wave Spectroscopy it is possible to measure the averaged
dynamical properties of a system through the dynamic absorption coefficient
$\mu(\omega,t)$ of Eq.\ (\ref{3.33}). Director modes are purely diffusive,
as described by Eq.\ (\ref{4.17}) for the structure factor, and possess
a viscosity $\eta_{\delta}(\bbox{q}_s)$ which we specify here \cite{Gennes93}:
\begin{mathletters}
\label{4.37}
\begin{eqnarray}
\eta_{1}(\bbox{q}_s) & = & \gamma -
 \frac{(\mu_3 q_{\perp}^2 - \mu_2 q_{\|}^2)^2}{\eta_b q_{\perp}^4 +
        \eta_{c} q_{\|}^4 + \eta_m q_{\perp}^2 q_{\|}^2} \\
\eta_{2}(\bbox{q}_s) & = & \gamma - 
 \frac{ \mu_2^2 q_{\|}^2}{\eta_a q_{\perp}^2 + \eta_b q_{\|}^2}
\end{eqnarray}
\end{mathletters}

\vspace*{.1cm}

with
\begin{mathletters}
\label{4.38}
\begin{eqnarray}
\eta_a & = & \mu_4 / 2 \\
\eta_b & = & (-\mu_2 + \mu_4 + \mu_5)/2 \\
\eta_c & = & (\mu_3 + \mu_4 + \mu_6)/2 \\
\eta_m & = & \mu_1 + \eta_b + \eta_c \enspace.
\end{eqnarray}
\end{mathletters}
The Leslie viscosities $\mu_i$ govern the viscous flow of the fluid and couple
it to the director motion. The Miesowicz viscosities $\eta_a$, $\eta_b$, and 
$\eta_c$ can be measured in pure flow experiments. The rotational viscosity 
$\gamma$ characterizes viscous forces due to rotations of the director.
With the explicit formula\ (\ref{4.17}) for the structure factor and small
enough times to expand the exponential time factor, $\mu(\omega,t)$ becomes
proportional to time $t$,
\begin{equation}
\label{4.39}
\mu(\omega,t) \approx (\Delta \varepsilon)^2 k_{\text{B}}T \, 
\frac{\pi^3 \omega^4}{2\overline{n^3} c^3} \, 
\sum_{\alpha,\beta,\delta} 
\int_{\hat{\bbox{k}}^{\alpha}}\int_{\hat{\bbox{q}}^{\beta}} 
\frac{N(\bbox{e}_{\alpha},\bbox{e}_{\beta},\hat{\bbox{u}}_{\delta})}{
       \eta_{\delta}(\bbox{q}_s)} \, t  \enspace.
\end{equation}
If we collect all the prefactors, the dynamic 
absorption coefficient can be written as
\begin{equation}
\label{4.39b}
\mu(\omega,t) = \mu_0 \, t \enspace \text{with} \enspace
\mu_0 = \alpha^2\,\frac{2k_{\text{B}}T}{9\pi} \, \frac{\omega^4}{c_{\perp}^3}
 \, \frac{\widetilde{\mu}}{\gamma} \enspace.
\end{equation}
Here the numerical factor $\widetilde{\mu}$ is a dimensionless angular
average involving the geometrical factor and the viscosities of the 
director modes. It depends on the Leslie viscosities relative to $\gamma$ and
the dielectric anisotropy $\alpha$, and it is equal 1 if 
$\eta_{\delta}(\bbox{q}_s) = \gamma$ and $\alpha=0$. When 
$\eta_{\delta}(\bbox{q}_s) = \gamma$, $\widetilde{\mu}$ can be evaluated
analytically even when $\alpha \ne 0$, and we find
$\widetilde{\mu} = \frac{1+\alpha/4}{1+\alpha/2}$. For the compound
5CB, $\gamma/\widetilde{\mu} = 0.60 \pm .20 \, \text{poise}$ was 
experimentally determined by Kao {\em et al.\/} using DWS and the last
formula \cite{Kao96,Jester96}. 
This value is in good agreement with the rotational viscosity
$\gamma = 0.81 \, \text{poise}$ of 5CB \cite{Collings95} and shows the 
validity of the theory. Of course $\gamma$ is larger than the measured
$\gamma/\widetilde{\mu}$ since $\eta_{\delta}(\bbox{q}_s)$ is smaller than
$\gamma$ [see Eqs.\ (4.37)] so that $\widetilde{\mu}$ exceeds 1.
However, the values of the Leslie viscosities are such that $\widetilde{\mu}$
is of order 1 in usual thermotropic nematics.
Furthermore, the Leslie-Erickson theory seems to describe the
director modes properly down to $4 \times 10^{-8}\,\text{s}$, the time 
resolution of the experiments, since there is no indication for a deviation 
from the linear time depencence of $\mu(\omega,t)$ predicted by the diffusive
director modes. Materials where the viscosities $\eta_{\delta}(\bbox{q}_s)$ 
for different values of $\bbox{q}_s)$ differ by factors of $10^2-10^3$
 are polymer nematic liquid crystals 
\cite{Taratuta85} with some director modes relaxing on a much larger 
time scale than they do in ordinary nematics. It would be interesting to 
study such systems to see if they exhibit
a deviation from the Leslie-Erickson theory for short times which, e.g., 
would show up in a different temporal power law for $\mu(\omega,t)$.

Finally, we point out an important difference between nematic liquid crystals
and colloidal suspensions. The dynamic absorption coefficient\ (\ref{4.39}) 
only 
contains the viscosities of the director modes. The Frank elastic constants 
cancel because they determine both the
light scattering and the dynamics. On the other hand, in colloidal suspensions
\cite{Weitz92} $\mu_0 = 2 D_{\text{B}} \omega^2/ (l^{\ast} c)$, where the
transport mean-free-path $l^{\ast}$ characterizes light propagation and the
diffusion constant $D_{\text{B}}$ the Brownian motion of the colloidal
particles.


\acknowledgments
We thank the Deutsche Forschungsgemeinschaft for financial support under
Grant No. Sta\ 352/2-2 and the NSF under Grant No. DMR 94-23114. We 
thank Ming Kao, Kristen Jester, and Arjun Yodh, who contributed to the
work by many valuable discussions.

\appendix

\section{The Bethe-Salpeter equation} \label{sec app1}

The averaged two-particle Green function with all space-time coordinates 
looks like
\begin{eqnarray}
\lefteqn{
\bbox{\Phi}(\bbox{x}_1,\bbox{y}_1,\tau_{1},\overline{\tau}_1;
 \bbox{x}_2,\bbox{y}_2,\tau_2,\overline{\tau}_2) \, = \, \qquad} \nonumber\\
\label{A1.1}
 & & \qquad
 \langle \, \bbox{G}^{R}(\bbox{x}_1,\bbox{x}_2;\tau_1,\tau_2) \otimes 
 \bbox{G}^{A}(\bbox{y}_2,\bbox{y}_1;\overline{\tau}_2,
\overline{\tau}_1)\, \rangle^{(N)} \enspace.
\end{eqnarray}
After an integration over electric field sources $\bbox{W}_0(2)$ at points 
$(\bbox{x}_2,\tau_2)$ and $(\bbox{y}_2,\overline{\tau}_2)$ it
provides us with the full autocorrelation function $\bbox{W}(1)$ of the 
electric light 
field at $(\bbox{x}_1,\tau_1)$ and $(\bbox{y}_1,\overline{\tau}_1)$:
\begin{equation}
\bbox{W}(1) = \int d2 \,\bbox{\Phi}(1,2) \bbox{W}_0(2) \enspace.
\end{equation}
In the following we abbreviate a pair of points in space time by its index
$i$. The two-particle Green function follows from the Bethe-Salpeter 
equation \cite{Frisch68}
\begin{equation}
\label{A1.2}
\bbox{\Phi}(1,4) = \bbox{f}(1,4) + 
\int d2 \, d3 \, \bbox{f}(1,2) \bbox{U}(2,3) \bbox{\Phi}(3,4) \enspace .
\end{equation}
The quantity $\bbox{f}(1,2)$ stands for the tensor product of the averaged 
one-particle Green functions:
\begin{eqnarray}
\bbox{f}(1,2) & = & 
[\langle \bbox{G}^{R} \rangle (\bbox{x}_1 -\bbox{x}_2;\tau_1 - \tau_2) 
\nonumber \\
\label{A1.3}
 & & 
\qquad \otimes \,  \langle \bbox{G}^{A} \rangle (\bbox{y}_2 -\bbox{y}_1;
\overline{\tau}_2 - \overline{\tau})]^{(N)} \enspace.
\end{eqnarray}
It propagates two electric field modes from $2$ to $1$ between their 
scattering events. The irreducible vertex function $\bbox{U}$ describes
different characteristic sets of scattering events \cite{Frisch68}. 
We will soon specify it.
Note, that there is no preferred point in time and that the scattering medium 
on average is homogeneous in space. Therefore, all our averaged quantities 
do not change under translations in space time, and hence they can only depend
on differences of the coordinates. We now introduce center-of-``mass'' 
($\bbox{R}_i,T_i$) and relative ($\bbox{r}_i,t_i$) coordinates:
\begin{equation}
\begin{array}{rcl@{\qquad}rcl}
\bbox{x}_i & = & \bbox{R}_i + \bbox{r}_i/2 & 
\bbox{y}_i & = & \bbox{R}_i - \bbox{r}_i/2 \\
\tau_i & = & T_i + t_i/2 & \overline{\tau}_i & = & T_i - t_i/2 \enspace.
\end{array}
\end{equation}
It is straightforward to show that the Jacobian determinant for this 
coordinate transformation is $1$, and we have
\begin{equation}
di = d^3R_i\,d^3r_i \, dT \, dt \enspace.
\end{equation}
All of our quantities only depend on differences in the center-of-``mass'' 
coordinates $\bbox{R}_i$ and $T_i$ because of the homogeneity in space time.
To discuss the Bethe-Salpeter equation we perform a Fourier transformation,
\begin{equation}
 \begin{array}{l}
  \int d^3(R_i-R_j)\,d^3(T_i-T_j)\,d^3r_i\,d^3r_j\,d^3t_i\,d^3t_j \\
  \quad \dots \, \text{exp}[-i(\bbox{K}\cdot(\bbox{R}_i - \bbox{R}_j)+ 
    \bbox{k}_i \cdot \bbox{r}_i - \bbox{k}_j \cdot \bbox{r}_j)] \\
  \enspace \qquad \times \, \text{exp}[i(\Omega (T_i - T_j)+ 
    \omega_i t_i - \omega_j t_j)] \enspace,
 \end{array}
\end{equation}
which transforms our quantities as follows:
\begin{equation}
 \begin{array}{rcl}
\bbox{\Phi}(1,4) & \,\longrightarrow \,& 
\bbox{\Phi}^{\omega_1\omega_4}_{\bbox{k}_1 
  \bbox{k}_{4\rule[-3mm]{0mm}{3mm} }}(\bbox{K},\Omega) \\
\bbox{U}(2,3) & \, \longrightarrow \, & 
\bbox{U}^{\omega_2\omega_3}_{\bbox{k}_2 
  \bbox{k}_{3 \rule[-2mm]{0mm}{2mm}}}(\bbox{K},\Omega) \\
\bbox{f}(1,2) & \, \longrightarrow \, & 
  \bbox{f}^{\omega_1}_{\bbox{k}_1}(\bbox{K},\Omega) \, 
  \delta_{\omega_1\omega_2} \, \bbox{1}^{(4)}_{\bbox{k}_1\bbox{k}_2} \enspace,
 \end{array}
\end{equation}
where $\delta_{\omega_1\omega_2}= 2\pi \, \delta(\omega_1 - \omega_2)$, and
$\bbox{1}^{(4)}_{\bbox{k}_1\bbox{k}_2}$, defined in Eq.\ (\ref{3.20}),
contains $\delta(\bbox{k}_1-\bbox{k}_2)$. The delta functions appear since
$\bbox{f}(1,2)$ only depends on the differences of the relative coordinates.
In Eq.\ (\ref{3.19}) we give the explicit form of 
$\bbox{f}^{\omega_1}_{\bbox{k}_1}(\bbox{K},\Omega)$.
The Fourier transformed Bethe-Salpeter equation finally takes the form
\begin{eqnarray}
&
\displaystyle
 \int \frac{d^3k_1}{(2\pi)^3}\frac{d\omega_1}{2\pi}\,\bigl[
 \bbox{1}_{\bbox{k}\bbox{k}_1}^{(4)} \delta_{\omega \omega_1} - 
 \bbox{f}^{\omega}_{\bbox{k}}(\bbox{K},\Omega)\,
 \bbox{U}^{\omega\omega_1}_{\bbox{k}\bbox{k}_1}(\bbox{K},\Omega) \bigr] 
  & \nonumber\\
\label{A1.8}
 & \qquad \times
 \bbox{\Phi}^{\omega_1\omega'}_{\bbox{k}_1\bbox{k}'}(\bbox{K},\Omega) 
= \bbox{f}^{\omega}_{\bbox{k}}(\bbox{K},\Omega) 
\bbox{1}_{\bbox{k}\bbox{k}'}^{(4)} \delta_{\omega \omega'} \enspace
\end{eqnarray}
and the autocorrelation function $\bbox{W}(\bbox{K},\bbox{k},\Omega,\omega)$
for light with frequency $\omega$ and wave vector $\bbox{k}$ follows from
\begin{eqnarray}
\bbox{W}(\bbox{K},\bbox{k},\Omega,\omega) & = & \int \frac{d^3k'}{(2\pi)^3}
\frac{d\omega'}{2\pi} \, 
\bbox{\Phi}^{\omega\omega'}_{\bbox{k}\bbox{k}'}(\bbox{K},\Omega) \nonumber\\
\label{A1.8a}
 & & \qquad \qquad \times \, 
 \bbox{W}_0(\bbox{K},\bbox{k}',\Omega,\omega') \enspace.
\end{eqnarray}

\begin{figure}
\centerline{\psfig{figure=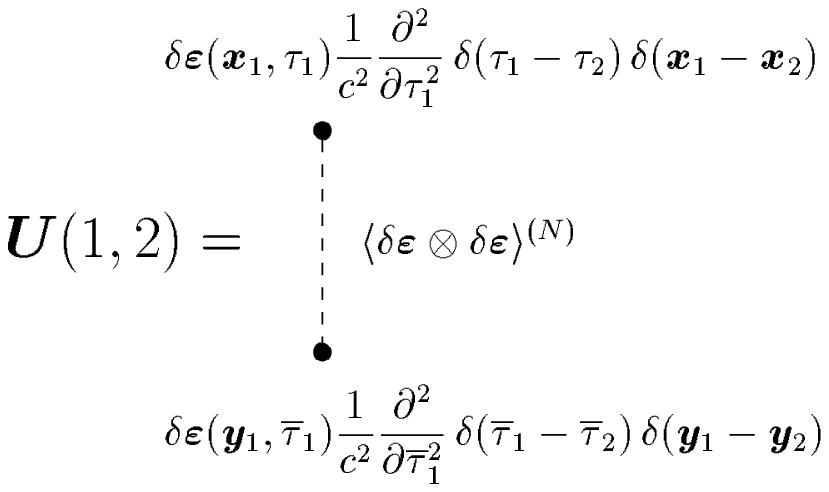}}

\vspace{.3cm}

\caption[]{The irreducible vertex function $\bbox{U}(1,2)$ in the 
weak-scattering  ap\-proxi\-ma\-tion. The second pair of points in
space time is tied to the structure factor via delta functions.}
\label{fig10}
\end{figure}

So far, our manipulations are generally valid. Now we introduce some
approximations. We use the weak-scattering approximation in which the
irreducible vertex function is given by \cite{Frisch68}
\begin{eqnarray}
\lefteqn{
\bbox{U}(1,2) \approx  \langle \, \delta \bbox{\varepsilon}(\bbox{x}_1,\tau_1)
\otimes \delta \bbox{\varepsilon}(\bbox{y}_1,\overline{\tau}_1) \, 
\rangle^{(N)} \, \frac{1}{c^4}\,\frac{\partial^2}{\partial \tau_1^2}\,
\frac{\partial^2}{\partial \overline{\tau}_1^2} } \quad \enspace \nonumber \\
 & & \times \, \delta(\bbox{x}_1-\bbox{x}_2)\delta(\bbox{y}_1-\bbox{y}_2)
\delta(\tau_1 - \tau_2) \delta(\overline{\tau}_1 - \overline{\tau}_2) \enspace.
\end{eqnarray}
It only considers scattering events of the two electric field modes which are
tied together through the structure factor. Fig.\ \ref{fig10} gives a 
graphic representation of $\bbox{U}(1,2)$. 
In center-of-``mass'' and 
relative coordinates we obtain
\begin{eqnarray}
\lefteqn{
\bbox{U}(1,2)  =  \langle \, \delta \bbox{\varepsilon}(\bbox{r}_1,t_1)
\otimes \delta \bbox{\varepsilon}(\bbox{0},0) \, 
\rangle^{(N)} \, \frac{1}{c^4}\,\frac{\partial^4}{\partial t_1^4}} \qquad
 \enspace \nonumber \\
 & & \textstyle \times \, 
    \delta(\bbox{R}_1-\bbox{R}_2 + \frac{\bbox{r}_1 - \bbox{r}_2}{2}) 
    \delta(\bbox{R}_1-\bbox{R}_2 - \frac{\bbox{r}_1 - \bbox{r}_2}{2}) \\
 & & \textstyle \times \, \delta(T_1-T_2 + \frac{t_1-t_2}{2}) 
         \delta(T_1-T_2 - \frac{t_1-t_2}{2}) \enspace , \nonumber
\end{eqnarray}
where we neglect the partial derivative $\partial / \partial T$ which probes
time variations on much longer time scales than the time period of light.
Then, in Fourier space, we obtain
\begin{equation}
\bbox{U}^{\omega_1\omega_2}_{\bbox{k}_1\bbox{k}_2}(\bbox{K},\Omega) =
\frac{\omega_2^4}{c^4} \,
\langle \delta \bbox{\varepsilon} \otimes \delta \bbox{\varepsilon} 
\rangle^{(N)}(\bbox{k}_1-\bbox{k}_2, \omega_1-\omega_2) \enspace,
\end{equation}
where the time derivative of a delta function was handled with its 
representation via plane waves. Since the temporal correlations in the
dielectric tensor decay on a time scale much longer than the time period
of light, $\bbox{U}^{\omega_1\omega_2}_{\bbox{k}_1\bbox{k}_2}(\bbox{K},\Omega)$
 is strongly peaked around $\omega_1=\omega_2$. The Bethe-Salpeter
equation (\ref{A1.8}) then implies the same behavior for
$\bbox{\Phi}^{\omega\omega'}_{\bbox{k}\bbox{k}'}(\bbox{K},\Omega)$, i.~e.,
$\bbox{\Phi}^{\omega\omega'}_{\bbox{k}\bbox{k}'}(\bbox{K},\Omega) \propto
p(\omega-\omega') g(\omega')$, where $p(\omega-\omega')$ stands for the 
strongly peaked part around $\omega=\omega'$ and $g(\omega')$ for 
the remaining smooth function in $\omega'$. The Green function 
$\bbox{\Phi}^{\omega\omega'}_{\bbox{k}\bbox{k}'}(\bbox{K},\Omega)$ picks up
a source term of frequency $\omega'$ and produces an autocorrelation function
with frequencies $\omega$ centered narrowly around $\omega'$. In the 
time domain this corresponds to $\text{exp}(i \omega' t)$ times a slowly
varying factor in $t$.
It is therefore appropriate to introduce the Green function
\begin{equation}
\bbox{\Phi}^{\omega'}_{\bbox{k} \bbox{k}'}(\bbox{K},\Omega,t) =
\int d^3(\omega-\omega') \, \bbox{\Phi}^{\omega\omega'}_{\bbox{k} \bbox{k}'}
\,e^{-i(\omega-\omega')t} \, ,
\end{equation}
which gives this factor for light sources of frequency $\omega'$. If we 
multiply Eq.\ (\ref{A1.8}) by $\text{exp}[-i(\omega-\omega')t]$, rewrite the
argument of $\langle \delta \bbox{\varepsilon} \otimes \delta 
\bbox{\varepsilon} \rangle^{(N)}$ as $\omega - \omega_1 = 
\omega-\omega'-(\omega_1-\omega')$, and integrate over $t$ we 
finally arrive at the Bethe-Salpeter equation of 
subsection\ \ref{subsec rand.twoG} for 
$\bbox{\Phi}^{\omega'}_{\bbox{k} \bbox{k}'}(\bbox{K},\Omega,t)$:
\begin{eqnarray}
\lefteqn{\displaystyle
 \int \frac{d^3k_1}{(2\pi)^3} [\bbox{1}_{\bbox{k}\bbox{k}_1}^{(4)} - 
\bbox{f}^{\omega'}_{\bbox{k}}(\bbox{K},\Omega)\,
\bbox{B}^{\omega'}_{\bbox{k}\bbox{k}_1}(t)]
\bbox{\Phi}^{\omega'}_{\bbox{k}_1\bbox{k}'}(\bbox{K},\Omega,t)} \nonumber\\
 & &\hspace*{4cm}  = \, \bbox{f}^{\omega'}_{\bbox{k}}(\bbox{K},\Omega) 
\bbox{1}_{\bbox{k}\bbox{k}'}^{(4)}
\end{eqnarray}
with
\begin{equation}
\bbox{B}^{\omega'}_{\bbox{k}\bbox{k}_1}(t) = 
\frac{(\omega')^4}{c^4} \,
\langle \delta \bbox{\varepsilon} \otimes \delta \bbox{\varepsilon} 
\rangle^{(N)}(\bbox{k}-\bbox{k}_1,t) \enspace.
\end{equation}
In deriving the last equation we replaced the argument $\omega$ in
$\bbox{f}^{\omega}_{\bbox{k}}(\bbox{K},\Omega)$ in Eq.\ (\ref{A1.8}) by
$\omega'$ and the factor $\omega_1^4$ in 
$\bbox{U}^{\omega\omega_1}_{\bbox{k}\bbox{k}_1}(\bbox{K},\Omega)$ by 
$(\omega')^4$.

\section{Two identities} \label{sec app2}
\subsection{Ward Identity} \label{subsec app2.ward}

The Ward identity establishes a linear relation between the irreducible 
vertex function $\bbox{U}$ and the mass operator.  In the weak-scattering
approximation it says
\begin{equation}
\label{A2.1}
\Delta \bbox{\Sigma}_{\bbox{k}}^{\omega}(\bbox{K} , 0 ) =
\int \frac{d^3 k'}{(2\pi)^3} \, \bbox{B}_{\bbox{k}\bbox{k}' }^{\omega}(t=0) \,
 \Delta \bbox{G}_{\bbox{k}'}^{\omega}(\bbox{K} , 0 )  \enspace,
\end{equation}
which can be proven by a variable transformation of the integrand.
For the general case see Vollhardt and W\"olfle \cite{Vollhardt80}.

\subsection{A useful identity} \label{subsec app2.use}

In this subsection we derive a useful relation between 
$\Delta \bbox{G}_{\bbox{k}}^{\omega}(\bbox{K} , \Omega) $, 
$\bbox{f}_{\bbox{k}}^{\omega}(\bbox{K} , \Omega)$, and 
$\Delta \bbox{\Sigma}_{\bbox{k}}^{\omega}(\bbox{K} , \Omega)$.
We start with the definition
\begin{equation}
\label{A2.2}
\Delta \bbox{G}_{\bbox{k}}^{\omega}(\bbox{K} , \Omega) 
= \langle \bbox{G}^{R} \rangle(\bbox{k}_{+},\omega_{+} )
  - \langle \bbox{G}^{A} \rangle(\bbox{k}_{-},\omega_{-} )
\end{equation}
and insert  the unit tensor $\bbox{1} $ in the appropriate representation:
\begin{eqnarray}
\lefteqn{ 
\Delta \bbox{G}_{\bbox{k}}^{\omega}(\bbox{K} , \Omega) 
\,= \,\langle \bbox{G}^{R} \rangle(\bbox{k}_{+},\omega_{+} ) 
 \, \big( [\langle \bbox{G}^{A} \rangle(\bbox{k}_{-},\omega_{-} )]^{-1} } 
\nonumber\\
\label{A2.2a}
    & &   \qquad \qquad \qquad
 - \,[\langle \bbox{G}^{R} \rangle(\bbox{k}_{+},\omega_{+} ) ]^{-1} \big) \,
   \langle \bbox{G}^{A} \rangle(\bbox{k}_{-},\omega_{-} ) \enspace.
\end{eqnarray}
Then we introduce  
$ \langle \bbox{G}^{R/A} \rangle(\bbox{k}_{\pm},\omega_{\pm}) $ from 
Eq.\ (\ref{3.9}), use  definitions (\ref{3.16b}) and (\ref{3.19})  for  
$\Delta \bbox{\Sigma}_{\bbox{k}}^{\omega}(\bbox{K} , \Omega)$ and
$\bbox{f}_{\bbox{k}}^{\omega}(\bbox{K} , \Omega)$, and arrive at the final equation:
\begin{eqnarray}
\lefteqn{
\Delta \bbox{G}_{\bbox{k}}^{\omega}(\bbox{K} , \Omega)  =
\bbox{f}_{\bbox{k}}^{\omega}(\bbox{K} , \Omega) \, 
[ \Delta \bbox{\Sigma}_{\bbox{k}}^{\omega}(\bbox{K} , \Omega) } \nonumber\\
\label{A2.3}
 & &  \qquad \qquad \qquad - \,\{ \bbox{G}_0^{-1}(\bbox{k}_{+} , \omega_{+})  -
         \bbox{G}_0^{-1}(\bbox{k}_{-} , \omega_{-})   \} ] \enspace.
\end{eqnarray}
In the case $K,\Omega \rightarrow 0$, we get to first order in $\bbox{K}$ and $\Omega$:
\begin{eqnarray}
\Delta \bbox{G}_{\bbox{k}}^{\omega}(\bbox{K} , \Omega)  & \approx &
\bbox{f}_{\bbox{k}}^{\omega}(\bbox{0} , 0) \, 
[ \Delta \bbox{\Sigma}_{\bbox{k}}^{\omega}(\bbox{0} ,  0) \nonumber \\
\label{A2.4}
 & & \qquad
 - \frac{\partial \bbox{G}_0^{-1} }{\partial \bbox{k} } \, \bbox{K} 
 - \frac{\partial \bbox{G}_0^{-1} }{\partial \omega} \, \Omega ] \enspace,
\end{eqnarray}
where we neglect first-order terms from $\bbox{f}_{\bbox{k}}^{\omega}$ and
$\Delta \bbox{\Sigma}_{\bbox{k}}^{\omega}$ because the components of 
$\Delta \bbox{\Sigma}_{\bbox{k}}^{\omega}$ are already much smaller than one.
The last equation shows that the leading order in $\bbox{K}$ and $\Omega$ 
comes solely from $\bbox{G}_0(\bbox{k},\omega)$.

In subsection \ref{subsec rand.oneG} we calculated the Green functions
$\langle \bbox{G}^{R/A} \rangle(\bbox{k},\omega)$ to first order in
$\bbox{\Sigma}^{R/A} (\bbox{k},\omega)$. They were diagonal and only involved
the diagonal elements of $\bbox{\Sigma}^{R/A} (\bbox{k},\omega)$.
Let us look at Eq.\ (\ref{A2.4}) under this approximation concentrating
on the propagating part of our quantities only. As explained in the main text
a Greek superscript or subscript corresponds, respectively, to the basis
vector $\bbox{e}_{\alpha}(\hat{\bbox{k}}) \otimes \bbox{e}_{\alpha}$ or
$\bbox{d}^{\alpha}(\hat{\bbox{k}}) \otimes \bbox{d}^{\alpha}$.
For $\bbox{f}_{\bbox{k}}^{\omega}(\bbox{0},0)$ we find
\begin{eqnarray}
\bbox{f}_{\bbox{k}}^{\omega}(\bbox{0} , 0)  & = & 
\bigl[\langle \bbox{G}^{R}\rangle (\bbox{k},\omega)  \otimes
\langle \bbox{G}^{A} \rangle (\bbox{k},\omega) \bigr]^{(N)} \nonumber\\
& \approx & \sum_{\alpha, \beta}
[\langle \bbox{G}^R\rangle(\bbox{k},\omega) ]^{\alpha}
[\langle \bbox{G}^A\rangle(\bbox{k},\omega) ]^{\beta}  \nonumber \\
\label{A2.6}
 & &
\qquad \bbox{e}_{\alpha}(\hat{\bbox{k}}) \otimes \bbox{e}_{\beta}
 (\hat{\bbox{k}})
\otimes \bbox{e}_{\alpha}(\hat{\bbox{k}}) \otimes \bbox{e}_{\beta}
 (\hat{\bbox{k}})
\enspace.
\end{eqnarray}
Terms like $\bbox{e}_1 \otimes\bbox{e}_2 \otimes \bbox{e}_1 \otimes \bbox{e}_2$
appear because the suberscript $(N)$ tells us to interchange the second and 
third basis vector in $\langle \bbox{G}^{R}\rangle  \otimes 
\langle \bbox{G}^{A} \rangle $. They only couple non-diagonal components to
each other. Since non-diagonal elements are beyond our approximation we don't
have to consider them. Then we are able to write
\begin{equation}
\label{A2.7}
[\bbox{f}_{\bbox{k}}^{\omega}(\bbox{0} , 0)]^{\alpha\beta} =
\left\{ \begin{array}{l}
                   [\langle \bbox{G}^R\rangle(\bbox{k},\omega) ]^{\alpha}
                   [\langle \bbox{G}^A\rangle(\bbox{k},\omega) ]^{\alpha} 
                   \enspace , \enspace \alpha=\beta \\
                   0 \enspace, \enspace \alpha \ne \beta
               \end{array}
\right.
\end{equation}
The component $0$ refers to terms like
$\bbox{e}_1 \otimes\bbox{e}_1 \otimes \bbox{e}_2 \otimes \bbox{e}_2$.
The derivation of $\bbox{G}_0^{-1}$ with respect to $\omega$ just gives
\begin{equation}
\label{A2.8}
\frac{\partial [\bbox{G}_0^{-1}]_{\alpha} }{\partial \omega} \, \Omega = 
\frac{2\omega}{c^2} \, \Omega \enspace .
\end{equation}
The derivation with respect to $\bbox{k}$ contains two contributions:
\begin{eqnarray}
\frac{\partial \bbox{G}_0^{-1} }{\partial \bbox{k}}  & = &
\sum_{\alpha} \Bigl[ \frac{\partial [\bbox{G}_0^{-1}]_{\alpha} }{\partial 
 \bbox{k}} \,
 \bbox{d}_{\alpha}(\hat{\bbox{k}}) \otimes \bbox{d}_{\alpha}(\hat{\bbox{k}})  
 \nonumber \\
\label{A2.9}
 & & +  [\bbox{G}_0^{-1}]_{\alpha} \, \frac{\partial}{\partial \bbox{k}}\, 
[ \bbox{d}^{\alpha}(\hat{\bbox{k}})  \otimes 
 \bbox{d}^{\alpha}(\hat{\bbox{k}})] \Bigr] \enspace .
\end{eqnarray}
The second one only produces non-diagonal elements which we do not have to 
consider. This statement seems to be obvious because a small rotation of 
$\bbox{k}$ rotates $\bbox{d}_{\alpha}(\hat{\bbox{k}})$. But our basis vectors 
are more general and we have to look at it more carefully. We have to show
that $ \bbox{e}_{i}(\hat{\bbox{k}}) \cdot \frac{\partial}{\partial k_j}\, 
\bbox{d}^{\,i}(\hat{\bbox{k}})$ is zero. With 
$\bbox{d}^{\,i} = \bbox{\varepsilon}_0 \bbox{e}_{i}$ and $\bbox{\varepsilon}_0$
being symmetric and independent of $\bbox{k}$ we can write
\begin{equation}
\label{A2.10}
\bbox{e}_{i}(\hat{\bbox{k}}) \cdot \frac{\partial}{\partial k_j}\, 
\bbox{d}^{\,i}(\hat{\bbox{k}}) = \bbox{d}^{\,i}(\hat{\bbox{k}}) \cdot 
\frac{\partial}{\partial k_j}\, \bbox{e}_{i}(\hat{\bbox{k}}) \enspace.
\end{equation}
From the biorthogonality relation it is clear that
\begin{equation}
\label{A2.11}
\bbox{d}^{\,i}(\hat{\bbox{k}}) \cdot 
\frac{\partial}{\partial k_j}\, \bbox{e}_{i}(\hat{\bbox{k}})
= - \bbox{e}_{i}(\hat{\bbox{k}}) \cdot \frac{\partial}{\partial k_j}\, 
\bbox{d}^{\,i}(\hat{\bbox{k}}) \enspace,
\end{equation}
which verifies the statement. We are now able to write down Eq.\ (\ref{A2.4})
within our approximation:
\begin{eqnarray}
[\Delta \bbox{G}_{\bbox{k}}^{\omega}(\bbox{K} , \Omega)]^{\alpha} & \approx &
[\bbox{f}_{\bbox{k}}^{\omega}(\bbox{0} , 0)]^{\alpha \alpha} \, 
\Bigl( [\Delta \bbox{\Sigma}_{\bbox{k}}^{\omega}(\bbox{0} ,  0)]_{\alpha} 
 \nonumber \\
\label{A2.12}
 & & \qquad
 - \frac{\partial [\bbox{G}_0^{-1}]_{\alpha} }{\partial \bbox{k} } \cdot 
 \bbox{K} - \frac{2\omega}{c^2} \, \Omega \Bigr) \enspace .
\end{eqnarray}

\newpage
\onecolumn
\widetext

\section{Structure factor} \label{sec struc}

We give the two important structure factors in scaled form and
in the coordinates $C_{\bbox{k}}$, $S_{\bbox{k}}$, $C_{\bbox{q}}$, 
$S_{\bbox{q}}$ and $\varphi$. From the notation it is clear whether they 
belong to an extraordinary or ordinary light ray. The parameters are $\alpha$,
$\overline{K}_1$, $\overline{K}_2$ and $h$.
\begin{mathletters}
\label{A3.1}
\begin{eqnarray}
[\widetilde{\bbox{B}}^{\omega}_{\bbox{k}^{1}\bbox{q}^{2}}(0)]_{12} & = &
\frac{1}{(1+\alpha)^2}\,\frac{S^2_{\bbox{k}}}{Q^2_{\perp}} \,
\left[ \frac{S^2_{\bbox{k}} \text{sin}^2\varphi}{\overline{K}_1 Q^2_{\perp}
+Q^2_{\|} + h^2} +
\frac{(S_{\bbox{k}} \text{cos}\varphi-S_{\bbox{q}})^2}{\overline{K}_2 
Q^2_{\perp} + Q^2_{\|} + h^2} \right] \\ \protect
[\widetilde{\bbox{B}}^{\omega}_{\bbox{k}^{1}\bbox{q}^{1}}(0)]_{11}  & = & 
\frac{1}{(1+\alpha)^2}\,\frac{1}{Q^2_{\perp}} \,
\left[
\frac{\text{cos}^2\varphi \, N_1 + 2\text{cos}\varphi \, N_2 + N_3}{
\overline{K}_1 Q^2_{\perp} + Q^2_{\|} + h^2} + 
\frac{\text{sin}^2 \varphi \, N_4}{\overline{K}_2 Q^2_{\perp} + Q^2_{\|} + h^2}
\right]
\end{eqnarray}
\end{mathletters}
with
\begin{equation}
\label{A3.2}
Q^2_{\|} = (C_{\bbox{k}} - C_{\bbox{q}})^2 \qquad , \qquad
Q^2_{\perp} = S_{\bbox{k}}^2 - 2 S_{\bbox{k}}S_{\bbox{q}} \, \text{cos}\varphi
+S_{\bbox{q}}^2
\end{equation}
and
\begin{equation}
\label{A3.3}
\begin{array}{rcl@{\qquad , \qquad}rcl}
N_1 & = & (S_{\bbox{k}}^2 C_{\bbox{q}} - S_{\bbox{q}}^2 C_{\bbox{k}})^2 &
N_2 & = & S_{\bbox{k}} S_{\bbox{q}}\, 
(S_{\bbox{k}}^2 C_{\bbox{q}} - S_{\bbox{q}}^2 C_{\bbox{k}})\,
(C_{\bbox{k}}-C_{\bbox{q}}) \\[.5ex]
N_3 & = & S_{\bbox{k}}^2 S_{\bbox{q}}^2 \, (C_{\bbox{k}}-C_{\bbox{q}})^2 &
N_4 & = & (S_{\bbox{k}}^2 C_{\bbox{q}} + S_{\bbox{q}}^2 C_{\bbox{k}})^2
\end{array}
\end{equation}
The integration over $\varphi$ gives:
\begin{mathletters}
\label{A3.4}
\begin{eqnarray}
\int [\widetilde{\bbox{B}}^{\omega}_{\bbox{k}^{1}\bbox{q}^{2}}(0)]_{12} \,
 d \varphi & = & \frac{\pi}{2}\,\frac{S_{\bbox{k}}^2}{S_{\bbox{q}}^2}\,
\left[ \frac{1}{\overline{K}_2} - \frac{1}{\overline{K}_1} +
 \frac{1}{C(h)}\,\left( \frac{I^{-1}(\overline{K}_1)}{\overline{K}_1}-
  \frac{I^{-1}(\overline{K}_2)}{\overline{K}_2}\right) + 4 S_{\bbox{q}}^2 
  I(\overline{K}_2) \right] \\[1ex]
\int \text{cos}\varphi \, 
[\widetilde{\bbox{B}}^{\omega}_{\bbox{k}^{1}\bbox{q}^{2}}(0)]_{12} \, d \varphi
 & = & \frac{\pi}{4}\,\frac{S_{\bbox{k}}}{S_{\bbox{q}}^3}\,
\left[ \frac{C(h)+2\overline{K}_2 (S_{\bbox{k}}^2-S_{\bbox{q}}^2)}{
\overline{K}_2^2} -
\frac{C(h)+2\overline{K}_1 (S_{\bbox{k}}^2+S_{\bbox{q}}^2)}{
\overline{K}_1^2} \right. \nonumber\\
 & & \left. \qquad \quad \, + \, \frac{1}{C(h)}\,\left( 
\frac{C(h)+\overline{K}_1 (S_{\bbox{k}}^2+S_{\bbox{q}}^2)}{\overline{K}_1^2
\, I(\overline{K}_1)}
-\frac{C(h)+\overline{K}_2 (S_{\bbox{k}}^2+S_{\bbox{q}}^2)}{\overline{K}_2^2
\, I(\overline{K}_2)} \right) \right.  \\
 & & \left. \qquad \quad \, + \, \frac{4S_{\bbox{q}}^2}{\overline{K}_2}\,
 [C(h)+\overline{K}_2
(S_{\bbox{k}}^2+S_{\bbox{q}}^2)] \, I(\overline{K}_2)\right] \nonumber\\[1ex]
\int [\widetilde{\bbox{B}}^{\omega}_{\bbox{k}^{1}\bbox{q}^{1}}(0)]_{11} \,
 d \varphi & = & \frac{\pi}{2} \, \left[ 
  \left( \frac{S_{\bbox{k}}^2}{S_{\bbox{q}}^2} \, C_{\bbox{q}}^2 +
         \frac{S_{\bbox{q}}^2}{S_{\bbox{k}}^2} \, C_{\bbox{k}}^2 \, \right) \,
  \left(\frac{1}{\overline{K}_1}- \frac{1}{\overline{K}_2}\right)
  -2C_{\bbox{k}} C_{\bbox{q}} \, 
  \left(\frac{1}{\overline{K}_1} + \frac{1}{\overline{K}_2}\right) \right.
 \nonumber\\
 & & \left. \qquad + \, \frac{1}{C(h)}\,
  \left( \frac{S_{\bbox{k}}}{S_{\bbox{q}}} \, C_{\bbox{q}} +
         \frac{S_{\bbox{q}}}{S_{\bbox{k}}} \, C_{\bbox{k}} \, \right)^2 \,
  \left( \frac{I^{-1}(\overline{K}_2)}{\overline{K}_2} -
         \frac{I^{-1}(\overline{K}_1)}{\overline{K}_1} \right) \right. \\
 & & \left . \qquad + \, \frac{4}{\overline{K}_1}\,I(\overline{K}_1) \, 
     \Bigl( \overline{K}_1 (S_{\bbox{k}}^2 C_{\bbox{q}} +
              S_{\bbox{q}}^2 C_{\bbox{k}}) \, (C_{\bbox{k}}+C_{\bbox{q}})
            + C(h) C_{\bbox{k}} C_{\bbox{q}} \Bigr) \right] \nonumber \\[1ex]
\int \text{cos}\varphi \, 
[\widetilde{\bbox{B}}^{\omega}_{\bbox{k}^{1}\bbox{q}^{1}}(0)]_{11} \, d \varphi
 & = & \frac{\pi}{4}\,\frac{1}{S_{\bbox{k}}^3 S_{\bbox{q}}^3} \,
\left[ \frac{1}{\overline{K}_1^2}\,
\Bigl( (S_{\bbox{k}}^2 C_{\bbox{q}} - S_{\bbox{q}}^2 C_{\bbox{k}})^2 C(h)
       + 2 \overline{K}_1 (S_{\bbox{k}}^4 C_{\bbox{q}}^2 - 
                           S_{\bbox{q}}^4 C_{\bbox{k}}^2) \,
            (S_{\bbox{k}}^2-S_{\bbox{q}}^2) \Bigr) \right. \nonumber \\
 & & \left. -\,\frac{1}{\overline{K}_2^2}\, 
            (S_{\bbox{k}}^2 C_{\bbox{q}} + S_{\bbox{q}}^2 C_{\bbox{k}})^2 \,
   \Bigl(C(h) + 2 \overline{K}_2 (S_{\bbox{k}}^2 + S_{\bbox{q}}^2) \Bigr)
\right. \nonumber \\
 & & \left. + \, \frac{1}{C(h)} \, 
  (S_{\bbox{k}}^2 C_{\bbox{q}} + S_{\bbox{q}}^2 C_{\bbox{k}})^2 \,
  \left(
  \frac{C(h)+\overline{K}_2 (S_{\bbox{k}}^2+S_{\bbox{q}}^2)}{\overline{K}_2^2
\, I(\overline{K}_2)}
-\frac{C(h)+\overline{K}_1 (S_{\bbox{k}}^2+S_{\bbox{q}}^2)}{\overline{K}_1^2
\, I(\overline{K}_1)} \right) \right. \\
& & \left. +\,4S_{\bbox{k}}^2  S_{\bbox{q}}^2 I(\overline{K}_1) \,
   \frac{C(h)+\overline{K}_1 (S_{\bbox{k}}^2+S_{\bbox{q}}^2)}{\overline{K}_1^2}
 \, \Bigl(\overline{K}_1 (S_{\bbox{k}}^2 C_{\bbox{q}} +
   S_{\bbox{q}}^2 C_{\bbox{k}}) \, (C_{\bbox{k}}+C_{\bbox{q}})
  + C(h) C_{\bbox{k}} C_{\bbox{q}}    \Bigr) \right] \nonumber
\end{eqnarray}
\end{mathletters}
with
\begin{equation}
\label{A3.5}
C(h) = (C_{\bbox{k}}-C_{\bbox{q}})^2 + h^2 \qquad \text{and} \qquad 
I(\overline{K}_i) = \Bigl\{ 
 \bigl[\overline{K}_i (S_{\bbox{k}}-S_{\bbox{q}})^2 + C(h)\bigr] \, 
 \bigl[\overline{K}_i (S_{\bbox{k}}+S_{\bbox{q}})^2 + C(h)\bigr] 
\Bigr\}^{-1/2}
\enspace.
\end{equation}

The structure factor 
$[\widetilde{\bbox{B}}^{\omega}_{\bbox{k}^{1}\bbox{q}^{1}}(0)]_{11}$ diverges
for $h=0$ and $\bbox{q}^{1} \rightarrow \bbox{k}^{1}$. We give 
$[\widetilde{\bbox{B}}^{\omega}_{\bbox{k}^{1}\bbox{q}^{1}}(0)]_{11}$
for small $\varphi$ and $\delta C=C_{\bbox{k}}-C_{\bbox{q}}$:
\begin{eqnarray}
\lefteqn{
[\widetilde{\bbox{B}}^{\omega}_{\bbox{k}^{1}\bbox{q}^{1}}(0)]_{11} \,= \,
\frac{1}{(1+\alpha)^2} \, \frac{4C_{\bbox{k}}^2}{(1+\alpha)^2 C_{\bbox{k}}^2 
 \delta C^2/S_{\bbox{k}}^2 + S_{\bbox{k}}^2 \varphi^2}} \nonumber\\[1ex]
& &\qquad \qquad \times  \left[\frac{(1+\alpha)^2 C_{\bbox{k}}^2 \delta C^2}{
 [\overline{K}_1(1+\alpha)^2 C_{\bbox{k}}^2 / S_{\bbox{k}}^2 +1] \delta C^2
 + \overline{K}_1 S_{\bbox{k}}^2 \varphi^2 + h^2}
 +\frac{S_{\bbox{k}}^2 C_{\bbox{k}}^2 \varphi^2}{
 [\overline{K}_2(1+\alpha)^2 C_{\bbox{k}}^2 / S_{\bbox{k}}^2 +1] \delta C^2
 + \overline{K}_2 S_{\bbox{k}}^2 \varphi^2 + h^2} \right] \, .
\end{eqnarray}


%


\end{document}